\documentclass[amsmath, amssymb, twocolumn, pra, superscriptaddress]{revtex4-2}

\usepackage{graphicx}
\usepackage{dcolumn}
\usepackage{bm}
\usepackage{bbm}
\usepackage{cases}
\usepackage{color} 
\usepackage[colorlinks,linkcolor=blue,anchorcolor=blue,citecolor=blue,filecolor=blue,menucolor=blue,runcolor=blue,urlcolor=blue,frenchlinks=blue]{hyperref}
\usepackage[mathscr]{eucal}
\usepackage{float}
\usepackage{enumitem}

\begin{document}
	
	\title{Fate of moir\'e flat bands for a weakly repulsive Bose-Einstein condensate in one-dimensional \(\mathcal{PT}\)-symmetric bichromatic optical lattices} 
	
	\author{Enhong Cheng}
	\email{ehcheng@m.scnu.edu.cn}
	\thanks{These authors contributed equally to this work.}
	\affiliation{School of Physics, South China Normal University, Guangzhou 510006, China}
	
	\author{Yu Tan}
	\thanks{These authors contributed equally to this work.}
	\affiliation{School of Physics, South China Normal University, Guangzhou 510006, China}
	
	\author{Yanzhen Xu}
	\affiliation{School of Physics, South China Normal University, Guangzhou 510006, China}
	
	\author{Li-Jun Lang}
	\email{ljlang@scun.edu.cn}
	\affiliation{School of Physics, South China Normal University, Guangzhou 510006, China}
	\affiliation{Guangdong Provincial Key Laboratory of Quantum Engineering and Quantum Materials, South China Normal University, Guangzhou 510006, China}
	\affiliation{Guangdong-Hong Kong Joint Laboratory of Quantum Matter, South China Normal University, Guangzhou 510006, China}

	\begin{abstract}
		One-dimensional (1D) superlattices provide one simplified platform for exploring moir\'e physics from a low-dimensional perspective, with the ratio of lattice constants playing a role analogous to the twist angle in two-dimensional bilayers. Here we propose a 1D $\mathcal{PT}$-symmetric bichromatic optical lattice for a weakly repulsive Bose-Einstein condensate and investigate how the interplay of dissipation and interaction impacts the lowest moir\'e flat band. 
		Without interaction, we find that the lowest-band flatness induced by commensurate ratios exhibits a parity-dependent response to the $\mathcal{PT}$-symmetric imaginary potential due to the distinct $\mathcal{PT}$ pairing mechanism for the energy spectrum. 
		For ratios with even denominators (i.e., even parities), the level attraction and thus the $\mathcal{PT}$-symmetry breaking occur within the lowest two bands, leading to a monotonic broadening of the lowest flat band, whereas odd denominators (i.e., odd parities) yield a nonmonotonic response due to the $\mathcal{PT}$-symmetry breaking within the second and the third lowest bands instead while the lowest band remains purely real.
		This parity-dependent phenomenon can be understood by the perturbation theory.
		Furthermore, by solving the Gross-Pitaevskii equation, we also find that although the weak repulsive interaction can broaden the moir\'e bands alone, the combined effects of interaction and imaginary potential also lead to parity-dependent behaviors. 
		For even parities, band flattening is consistently diminished, whereas for odd parities, the imaginary potential can either enhance or reduce the degree of flattening. 
		These results pave the way for experimental studies of dissipation and interaction effects on band flatness in moir\'e systems.
	\end{abstract}
	\maketitle
	\section{Introduction \label{SEC.I}}
	Ultracold atoms are an ideal platform for quantum simulation, enabling the realization of complex lattice structures via highly controllable lasers that are difficult to fabricate in solid-state materials \cite{LewensteinSen2007AiP,BlochZwerger2008RMP,ZhangZhu2018AiP}. 	
	Recently a two-dimensional (2D) moir\'e model was implemented in a Bose-Einstein condensate (BEC) with spin-dependent optical lattices \cite{MengZhang2023Nature}. 
	This advance allows a reexamination of many-body effects in moir\'e physics \cite{CaoJarillo-Herrero2018Nature,CaoKaxiras2018Nature,LuDmitri2019nature,Yankowitz2019Science,JinKin2021nature} within a clean and highly tunable atomic setting. 
	To explore the physical essence of such effects in a more tractable setting, theoretical interest has naturally turned to a one-dimensional (1D) analog. 
	The 1D moir\'e lattice, a commensurate bichromatic lattice with a large unit cell, has been theoretically shown to host strongly flattened bands and interaction-driven correlated states similar to those in twisted 2D systems \cite{Vu2021}. 
	This has inspired studies on the interplay between interactions and (in)commensurability in 1D moir\'e systems \cite{Goncalves2024NP,ZhangZhang2025CommPhy}, and has also prompted its implementation in other platforms, such as photonics \cite{NguyenNguyen2022PRR,TalukdarRyckman2022ACSPho,YuChen2023PRL,XiaLiang2024OL,TrushinOta2025OL,LiChen2025PRL}. 
	Meanwhile, theoretical investigations of related phenomena have been conducted in cold-atom systems \cite{NathRoy2014LPL,NathRoy2022TEPJD,RaghavRoy2022PRA,ZhouZhang2025PRA}. 
	Due to extensive research on incommensurate bichromatic lattices \cite{Roati2008Roati,Schreiber2015Science,KohertAidelsburger2019PRL}, the experimental realization of the 1D moir\'e lattice is now well established with ultracold atoms.
	
	On the other hand, among the various research directions in non-Hermitian systems \cite{AshidaUeda2020,BergholtzKunst2021}, parity-time-reversal ($\mathcal{PT}$) symmetric systems constitute a crucial branch of study due to their potential to exhibit entirely real eigenvalue spectra \cite{BenderBoettcher1998PRL,Bender2007,KonotopDmitry2016RMP}. 
	Early investigations focused on $\mathcal{PT}$-symmetric optical waveguides with periodic modulation of complex refractive indices, where novel wave-dynamic phenomena such as Bloch oscillations were observed \cite{MusslimaniChristodoulides2008PRL,Longhi2009PRL}. 
	These works prompted further exploration into more generalized forms of periodic $\mathcal{PT}$-symmetric lattices. 
	In the linear regime, properties such as symmetry-breaking phase transitions and associated dynamics in models like the sinusoidal $\mathcal{PT}$-symmetric optical lattice have been extensively studied \cite{Makris2010,GraefeJones2011PRA,Jones2014}. 
	When nonlinearity is present, for example in the context of Kerr nonlinearity in optical waveguides or the Gross-Pitaevskii equation (GPE) describing mean-field BECs, a rich variety of nonlinear localized states and excitations can emerge in such lattices \cite{KonotopDmitry2016RMP}. 
	These include periodic Bloch waves \cite{AbdullaevYulin2010PRE} and various types of solitons \cite{ZhouLiu2010OL,ZhuHe2011OL,NixonYang2012OL,NixonYang2012PRA}. 
	Particularly noteworthy is the recent discovery of asymmetric swallowtail structures in nonlinear sinusoidal $\mathcal{PT}$-symmetric lattices \cite{ZhangKonotop2021PRL}, which highlights the significant impact of the interplay between nonlinearity and $\mathcal{PT}$ symmetry on the band structure and dynamical behavior of the system.
	
	Inspired by the correlated states in strongly flat bands of 1D moir\'e lattices~\cite{Vu2021} and the pronounced band-structure control in nonlinear $\mathcal{PT}$-symmetric systems~\cite{ZhangKonotop2021PRL}, it is natural to explore their combined effects. 
	A central question is how $\mathcal{PT}$ symmetry influences the flatness of the lowest moir\'e band, and how weak repulsive interactions interplay with non-Hermiticity to further modulate this flatness. 
	In this paper, we numerically study the lowest band of a BEC in a 1D $\mathcal{PT}$-symmetric moir\'e optical lattice using the GPE. 
	In the noninteracting regime, the parity-dependent response originates from distinct $\mathcal{PT}$-pairing structures within each moir\'e unit cell. 
	For ratios with even denominators (i.e., even parities), the lowest two bands couple directly through the imaginary potential, leading to monotonic broadening; 
	while for odd denominators (i.e., odd parities), a self-conjugate center region delays the $\mathcal{PT}$-symmetry breaking of the lowest band, and its competition with non-Hermitian confinement produces a nonmonotonic response. 
	This parity-dependent phenomenon can be understood by the perturbation theory.
	With weak repulsive interactions, the overall band dispersion increases, yet the combination of interaction and non-Hermiticity selectively enhances or suppresses flattening depending on the parity of ratio denominators. These results advance the understanding of how non-Hermitian effects and weak interactions cooperate or compete in modulating moir\'e band flatness.
	
	\section{GPE with 1D \(\mathcal{PT}\)-symmetric complex moir\'e potential}\label{SEC.II}
	We consider a BEC with a large particle number in a 3D potential composed of a transverse harmonic trap in the $(x,y)$ plane and a longitudinal optical lattice along the $z$ direction.
	This system can be modeled at the mean-field level by the 3D GPE, and the effective interaction between two atoms is described by $g_{\rm 3D} = 4\pi\hbar^2a_s/m$, where $a_s$ is the $s$-wave scattering length.
	Under sufficiently strong harmonic confinement such that no transverse excitations are induced by weak interactions, i.e., when $a_s^2/a_\perp^2\ll a_s n_z\ll1$, with $a_\perp$ and $n_z$ being the oscillator length and the average 1D density, the BEC can be effectively described by a reduced 1D GPE along the lattice direction \cite{SalasnichReatto2002PRA,LiebSeiringer2003PRL},
	\begin{equation}
		i \hbar \frac{\partial}{\partial t}\Psi(z,t) = \Big[ -\frac{\hbar^2}{2m}\frac{\partial^2}{\partial z^2} + V_{\rm OL}(z) + g |\Psi|^2 \Big]\Psi(z,t),
		\label{1DGPEtime}
	\end{equation}
	where $\Psi(z,t)$ is the 1D macroscopic condensate wave function, normalized over the system length $L$ to the total particle number $N$ such that $\int_L|\Psi(z,t)|^2 dz = N$. 
	Here, $m$ is the atomic mass, and $g = g_{\rm 3D}/2\pi a^2_\perp$ denotes the effective 1D interaction strength \cite{SalasnichReatto2002PRA}.
	The external potential
	\begin{equation}\label{OL}
		V_{\rm OL}(z) = V_0 \Big[ \cos^2(k_1 z) + \cos^2(k_2 z)+ i\frac{\gamma}{2}\sin(2k_2 z) \Big],
	\end{equation}
	represents a 1D complex bichromatic optical lattice (i.e., moir\'e lattice) with strength $V_0$,  where the real part is formed by superimposing a primary lattice with wave number $k_1$ and a secondary lattice with wave number $k_2$, and the imaginary part can be engineered by the atomic gain and loss controlled by the parameter $\gamma$. 
	Although the potential individually breaks parity $\mathcal{P}$ and time-reversal $\mathcal{T}$ symmetries, it satisfies the combined $\mathcal{PT}$ symmetry, i.e., $V_{\rm OL}(z) = V^*_{\rm OL}(-z)$.
	Since the potential is complex, the total particle number $N$ may change over time, corresponding to a change in the norm of the wave function \cite{Pethick2008PethickCU,HaagWunner2014PRA,GutohrleinWunner2015JoPA}.
	
	For convenience, we make the following substitution
	\begin{eqnarray}
		t\to  \dfrac{\hbar}{E_r} t ,~~~~~	z\to  \dfrac{1}{2k_1}z,~~~~~V_0\to 2{E_r}V_0,
	\end{eqnarray}
	where the recoil energy $E_r=\hbar^2 k_1^2/2m$ and the primary lattice constant divided by $2\pi$ are chosen as the energy and length units, respectively, to reduce Eq. \eqref{1DGPEtime} to the dimensionless form
	\begin{eqnarray}
		i \dfrac{\partial}{\partial t}\Psi = \Big[ -4\dfrac{\partial^2}{\partial z^2} + V(z) + c |\Psi|^2 \Big]\Psi,
		\label{1DGPEtimeDim}
	\end{eqnarray}
	with  $c = g/E_r$. The lattice potential becomes
	\begin{eqnarray}\label{lattice}
		V(z) =  V_0 \left[ \cos\left(z\right) + \cos\left(\alpha z\right)
		+ i\gamma\sin\left(\alpha z\right) \right], ~~~
	\end{eqnarray}
	where an overall constant $V_0$ has been omitted. 
	The parameter $\alpha \equiv  k_2/k_1$ is defined as the 1D moir\'e ratio \cite{Vu2021}, which gives rise to either aperiodic (incommensurate) or periodic (commensurate) potential functions. 
	Notable phenomena such as Anderson localization \cite{LyeInguscio2007PRA} and mobility edges \cite{Biddle2010Biddle,ZezyulinGeorgy2024PRA} have been studied in 1D incommensurate lattice models.
	In this work we focus on the commensurate case, where $\alpha = 1/q \in \mathbb{Q}$,  with denominator $q\ge1$. 
	The unit cell sizes of the primary and secondary lattices are $a_1 = 2\pi$ and $a_2 = 2\pi/\alpha$, respectively. The resulting commensurate potential $V(z)$ exhibits a period of $A = a_2 = q a_1 \ge a_1$, referred to as the 1D moir\'e cell \cite{Vu2021}.  
	
	To analyze the properties of the energy bands, we consider solutions of Eq. \eqref{1DGPEtimeDim} in the form of a stationary Bloch-type wave function
	\begin{eqnarray}
		\Psi_{k}(z,t) = e^{i \left(kz - \mu_k t\right)} \psi_{k}(z),
		\label{BlochStationary}
	\end{eqnarray}
	where $\psi_{k}(z)$ is the periodic wave function with period $A$, and $\mu_k$ is a chemical-potential-like parameter.
	The moir\'{e} Bloch wave vector $k$ is restricted to the first Brillouin zone $k \in [-K/2, K/2)$ with $K = 2\pi/A$.
	Other solutions to Eq. \eqref{1DGPEtimeDim} such as period-multiplied \cite{MachholmSmith2004PRAdoubling} and soliton \cite{ZhangWu2009PRL} solutions may also exist.
	For the real case of $\mu$, substituting  Eq. \eqref{BlochStationary} into Eq. \eqref{1DGPEtimeDim}, we find that $\psi_{k}(z)$ satisfies the time-independent GPE
	\begin{align}
		\mu_k \psi_k 
		&= \Big[ -4\Big(\dfrac{\partial}{\partial t} + ik\Big)^2 + V(z) + c|\psi_k|^2 \Big] \psi_k \label{timeIndepA} \\
		&\equiv H[\psi_k] \psi_k . \label{timeIndepB}
	\end{align}
	For non-Hermitian systems, we can expect the solutions in Eq. \eqref{BlochStationary} with complex $\mu$ may also exist, but it does not satisfy Eq. \eqref{timeIndepA} any more except for the non-interacting case (i.e., $c=0$), where the $\mu_k$ is reduced to the complex energy of the system.
	
	Given that the size of the BEC system is much larger than the moir\'e cell ($L \gg A$), the average particle number per cell can be expressed in terms of the average 1D density as $An_z$. 
	The wave function can be normalized as  
	\begin{eqnarray}
		\frac{1}{A}\int_{-A/2}^{A/2} dz |\psi_{ k}(z)|^2 = n_z, 
		\label{normpsi}
	\end{eqnarray}
	and the energy functional density is given by
	\begin{align}
		\varepsilon[\psi_k]
		&= \frac{1}{A}\int_{-A/2}^{A/2} dz \, \psi_k^*\Bigl( H[\psi_k] - \frac{c}{2}|\psi_k|^2 \Bigr)\psi_k \label{varepsilonA}\\
		&= \mu_k - \frac{1}{A}\int_{-A/2}^{A/2} dz \, \frac{c}{2}|\psi_k|^4  . \label{varepsilonB}
	\end{align}
	We define the energy bands by the energy per particle, $E_k \equiv \varepsilon[\psi_k]/n_z$, expressed as a function of $k$. 
	Here, the average particle density $n_z$ is independent of $k$ and is treated as a system parameter. 
	Similarly, for each moir\'e ratio, we fix $n_z$ such that the dimensionless $n_zc$ can be used to describe the interaction strength.
	
	For non-Hermitian systems, the energy is generally complex and the imaginary parts $E^{\rm (i)}$ may cause the dynamical instability. 
	Therefore, in the following we mainly pay attention to the real-energy case unless otherwise mentioned, which, according to Eq. \eqref{varepsilonA}, also corresponds to the real $\mu$ required by Eq. \eqref{timeIndepA} for $c\ne 0$.
	For simplicity, we use superscripts (r) and (i) to denote the real and imaginary parts of the corresponding quantities, respectively.
	
	\begin{figure}[tb]	
		\includegraphics[width=1\linewidth]{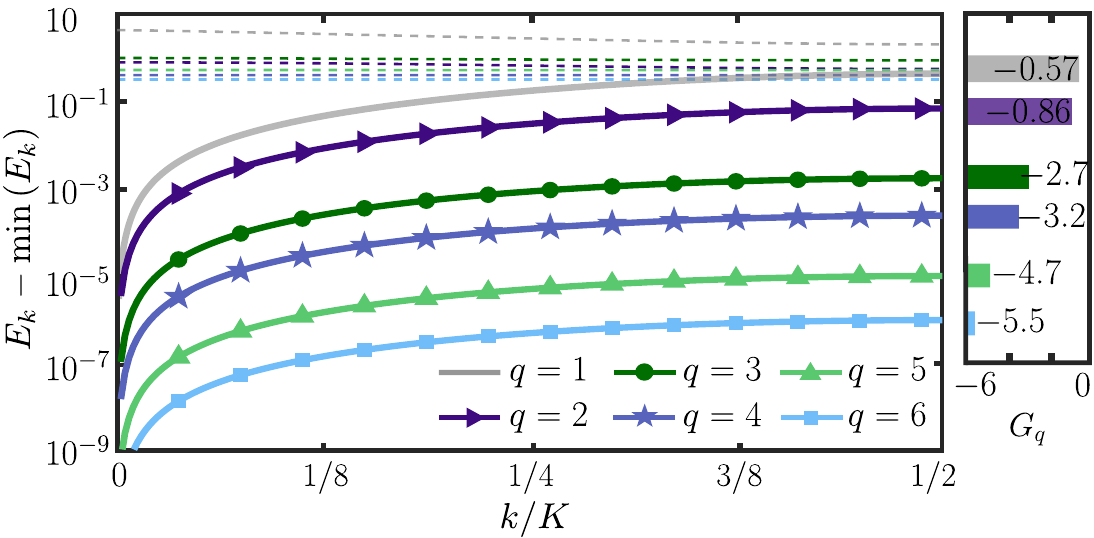}
		\caption{Lowest ($E_{\nu=1}$, solid lines) and second ($E_{\nu=2}$, dashed lines) bands of the non-interacting Hermitian system for moir\'e ratios $\alpha=1/q$ at $V_0=0.8$. The right panel shows the corresponding band gap ratio $G_q$ defined in Eq.~\eqref{gq}.}\label{fig1}
	\end{figure}
	
	\section {Non-Hermitian effect on moir\'e flat bands without interaction}\label{Sec.linear}
	\begin{figure*}[th]	
		\includegraphics[width=1\linewidth]{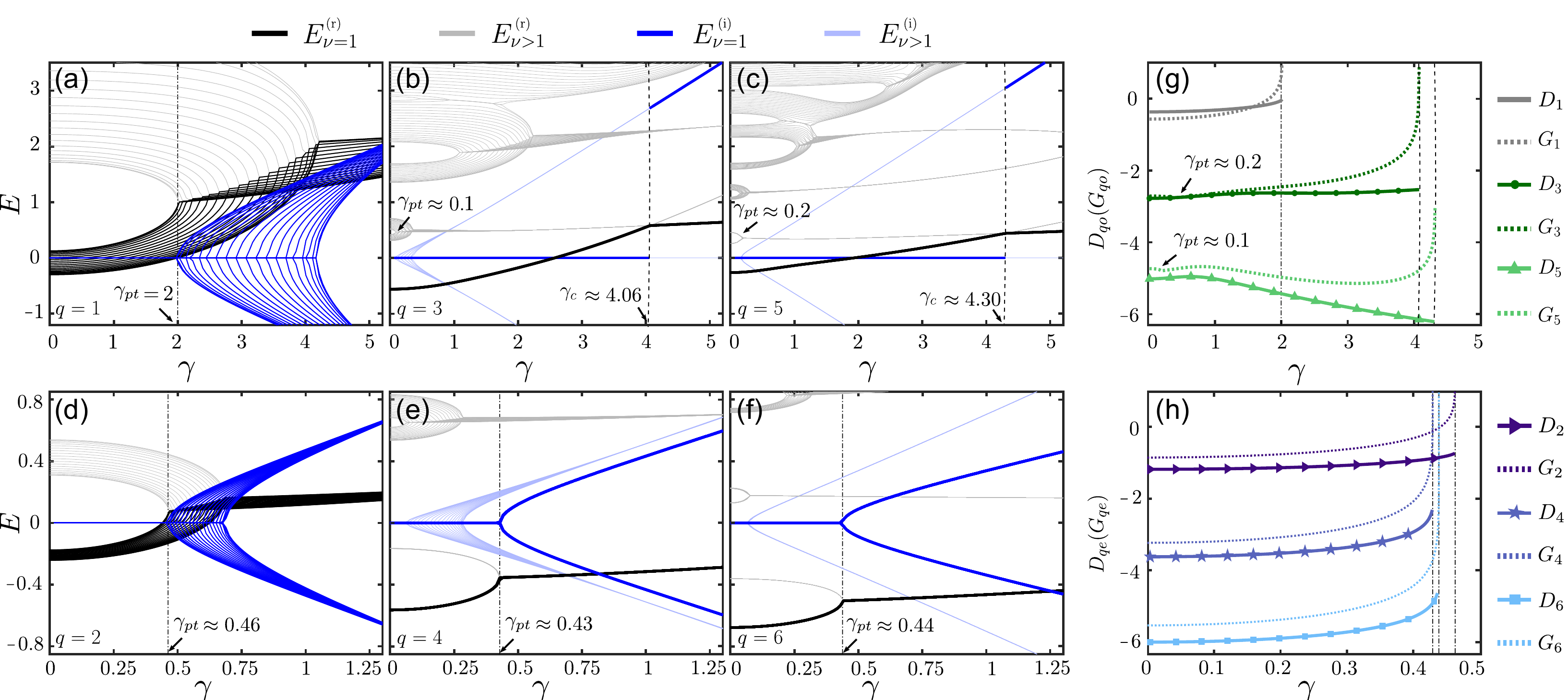}
		\caption{(a)-(f) Several lowest energy bands as a function of the non-Hermitian strength $\gamma$ for different moir\'e ratios $(1/q)$ in the absence of interactions. 
			The black and blue curves represent the real and imaginary parts of the lowest band ($E_{\nu=1}$), respectively; the gray curves represent the real parts of higher bands ($E_{\nu>1}$), and the light-blue lines correspond to the imaginary parts of a few selected higher bands.
			(a) Single-lattice $q=1$ (b)-(c) odd parities $q_{o} = \{3,5\}$, with the dispersion $D_{q}$ and gap ratio $G_{q}$ of the lowest real band plotted against $\gamma$ in (g) shown as solid and dashed lines, respectively; (d)-(f) even parities $q_{e}= \{2,4,6\}$, with the corresponding $D_{q_e}$ and $G_{q_e}$ shown in (h).
			The $\mathcal{PT}$-symmetry breaking points $\gamma_{pt}$ of the lowest band are determined numerically, marked by vertical dash-dotted lines in (a) and (d)-(f), and the real‑part crossing points $\gamma_c$ in (b) and (c) are marked by vertical dashed lines.}\label{fig2}
	\end{figure*}
	In this section, we consider the effects of the moir\'e ratios and non-Hermiticity on the lowest band and its flatness in the absence of interactions. 
	In the following analysis, we set the potential strength $V_0$ of the moir\'e lattice [Eq. \eqref{lattice}] to $0.8$.
	We consider a set of moir\'e ratios $\alpha = \{1/2, 1/3, 1/4, 1/5, 1/6\}$ and compare them with the single-lattice case of double depth at $\alpha = 1$. 
	These ratios correspond to configurations where the period of the secondary lattice is an integer multiple $q$ of the primary lattice period, yielding a moir\'e cell size of $A = 2\pi q$ for ratio denominators $q = \{1, 2, 3, 4, 5, 6\}$. 
	
	To characterize the flatness of the lowest band, we define the band dispersion ratio
	\begin{eqnarray}
		D_q = \log_{10} \left(w_q\right),
		\label{fq}
	\end{eqnarray}
	where $w_q \equiv \max(E_{\nu=1}) - \min(E_{\nu=1})$ is the bandwidth of the lowest real band.
	The integer band index $\nu\ge1$ is ordered by the real part of energy.
	To quantify the interplay between dispersion and excitation, we introduce the inverse gap ratio
	\begin{eqnarray}
		G_q = \log_{10} \left(w_q/\Delta_q\right),
		\label{gq}
	\end{eqnarray}
	where $\Delta_q \equiv \min(E^{\rm{(r)}}_{\nu=2}) - \max(E_{\nu=1})$ denotes the real band gap.  
	A value of $D_q (G_q)\to -\infty$ corresponds to a completely dispersionless lowest band, whereas $G_q \to +\infty$ indicates that the lowest band becomes gapless or its real part becomes degenerate.
	
	The lowest two bands for the Hermitian cases are shown in the main panel of Fig. \ref{fig1} as solid and dashed lines, respectively (see Appendix \ref{APPNumerical} for numerical details).
	As $q$ increases, the bandwidth of the lowest band ($E_{\nu=1}$) decays exponentially, consistent with Ref. \cite{Vu2021}. 
	In contrast, the band gap exhibits an overall oscillatory decrease, while decreasing monotonically when the even and odd denominators  are considered separately.
	Meanwhile, the gap ratio $G_q$ decreases with increasing $q$, as shown in the right subpanel of Fig. \ref{fig1}, demonstrating that the expansion of the moir\'e unit cell suppresses the band dispersion more effectively than it narrows the band gap.

	Figure~\ref{fig2} shows the noninteracting lowest complex energy bands, along with $D_q$ and $G_q$, as functions of $\gamma$. 
	The real (imaginary) parts of $E_{\nu=1}$ and $E_{\nu>1}$ are shown in black and gray (blue and light blue), respectively.
	For the single-lattice case [$q=1$, Fig.~\ref{fig2}(a)], as $\gamma$ increases, the bandwidth of $E_{\nu=1}$ widens monotonically, while the band gap narrows. 
	This trend is reflected in $D_{1}$ and $G_{1}$ in Fig.~\ref{fig2}(g) (solid and dashed lines, respectively). 
	$\mathcal{PT}$ symmetry breaks at $\gamma_{pt}=2$, where the lowest two bands coalesce at exceptional points (vertical dash-dotted line).
	This critical point follows from the transformation $z \to z - i\tanh^{-1}(\gamma/2)$ applied to Eq.~\eqref{lattice} \cite{Jones2014}.
	The even moir\'e parities $q_e = \{2,4,6\}$ exhibit similar behavior. 
	As shown in Figs.~\ref{fig2}(d)--\ref{fig2}(f), the first $\mathcal{PT}$-symmetry breaking also involves the lowest two bands, with numerically determined thresholds $\gamma_{pt}$.
	As $\gamma$ approaches $\gamma_{pt}$, the bandwidth of $E_{\nu=1}$ increases while the gap between the lowest two bands decreases; both $D_{q_e}$ and $G_{q_e}$ increase monotonically, with $G_{q_e}$ diverging at the EP in Fig.~\ref{fig2}(h).
	
	Odd denominators $q_{o} = \{3,5\}$ exhibit qualitatively different behavior.
	In contrast to the even case, the first $\mathcal{PT}$-breaking transition occurs between $E_{\nu=2}$ and $E_{\nu=3}$ at $\gamma_{pt}$, as illustrated in Figs.~\ref{fig2}(b) and \ref{fig2}(c).
	Beyond $\gamma_{pt}$, the two levels split into a complex-conjugate pair, while $E_{\nu=1}$ remains purely real and shifts upward.
	Meanwhile, the common real part of $E_{\nu=2,3}$ changes only weakly. 
	Over a broad range of $\gamma$ before the real-part crossing, $E_{\nu=1}$ stays purely real, with $D_{q_o}$ and $G_{q_o}$ in Fig.~\ref{fig2}(g) behaving nonmonotonically.
	At $\gamma_c$ in Figs.~\ref{fig2}(b) and \ref{fig2}(c), the real part of $E_{\nu=1}$ crosses the common real part of $E_{\nu=2,3}$, as marked by the vertical dashed lines. 
	This crossing reorders the bands by their real parts; the originally purely real band is no longer the lowest and is relabeled to a higher $\nu$, while the complex-conjugate pair is labeled $E_{\nu=1,2}$ in the real-part ordering.
	
\begin{figure}[tb]	
		\includegraphics[width=1\linewidth]{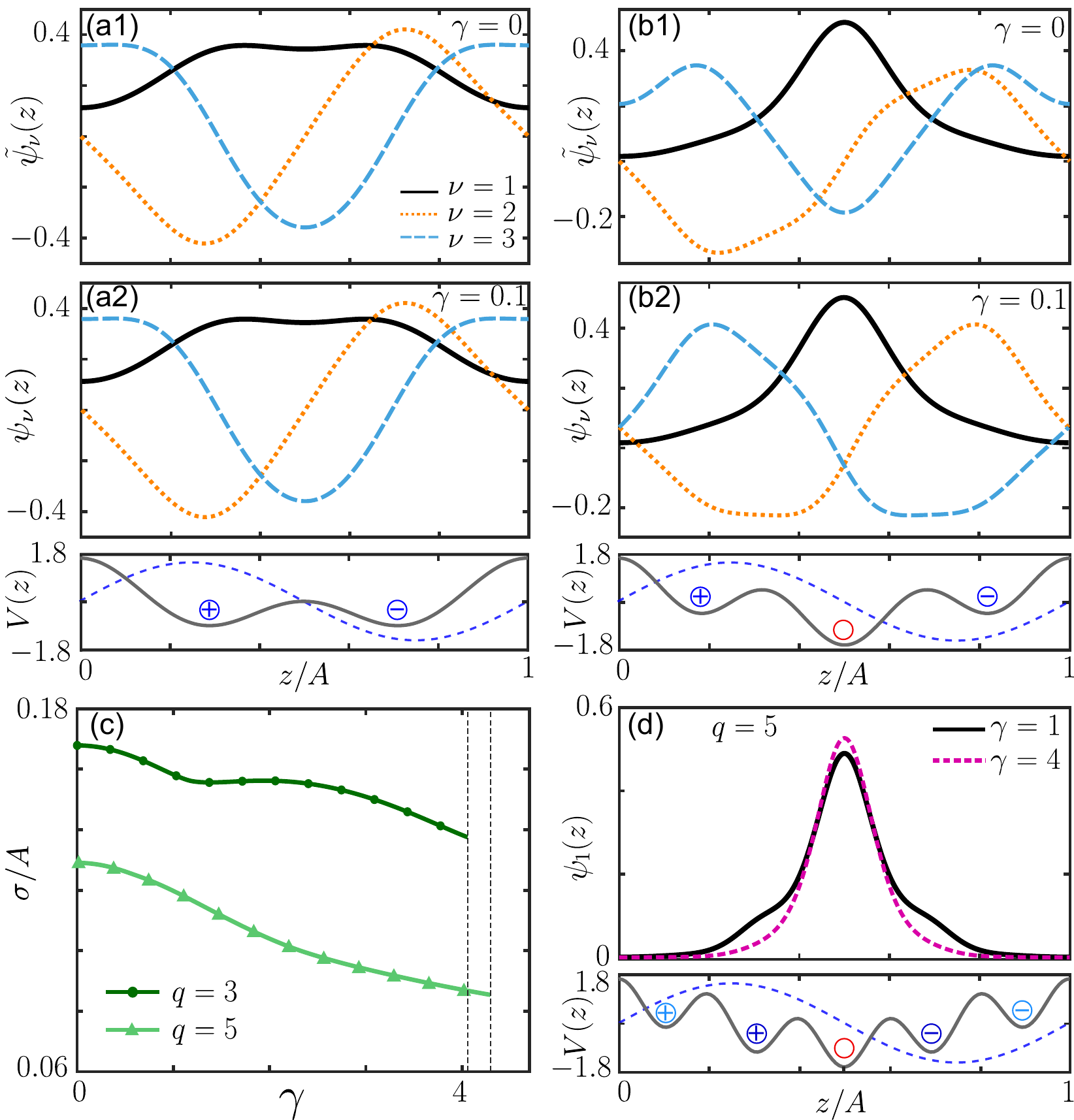}
		\caption{Wave functions $\psi_\nu(z)$ at $k=0$ for the lowest three bands (upper panels) and the corresponding potential profiles (lower panels) within one moir\'e unit cell for (a) $q=2$ and (b) $q=3$.
			The real and imaginary parts of the potential are shown by gray solid and blue dashed curves, respectively. 
			The circles mark minima of the potential real part from  Eq.~\eqref{lattice}, with the $\pm$ signs labeling the sign of its imaginary part; the red open circle specifically marks the center minimum, at which $V^{\rm (i)}(A/2)=0$.
			(a1) and (b1) represent $\tilde{\psi}_{\nu}$ in the Hermitian case, with $|\mathcal{V}_{21}|/|\mathcal{V}_{32}| \approx 3$ and $0.63$, respectively; 
			the corresponding non-Hermitian counterparts ($\gamma = 0.1$) in (a2) and (b2), where in the latter $\psi_2$ and $\psi_3$ are near an EP, with $\bigl|\frac{1}{A}\int_A \psi_3^*\,\psi_2\,dz\bigr| \approx 0.8$.
			(c) The normalized spatial width $\sigma/A$ of $\psi_1$ varies with $\gamma$.
			(d) Profiles of $\psi_1$ for $q=5$ at $\gamma = 1$ and $4$.
		}\label{fig3}
	\end{figure}

	To understand the parity-dependent flattening behaviors induced by this $\mathcal{PT}$-symmetric imaginary potential, we adopt the perturbation theory regarding small $\gamma$.
	The energy up to the second order in $\gamma$ yields
	\begin{eqnarray}
		E_\nu = \tilde{E}_\nu
		+ \gamma^2\sum_{\mu\ne\nu}
		\frac{|\mathcal{V}_{\mu\nu}|^2}
		{\Delta_{\mu\nu}}
		+ O(\gamma^3),
		\label{pertE}
	\end{eqnarray}
	where $\tilde{\psi}$ is the unperturbed wave function for Hermitian systems, and $\mathcal{V}_{\mu\nu} \equiv \frac{iV_0}{A} \int_A \tilde{\psi}_\mu^* V^{\rm (i)}\tilde{\psi}_\nu dz = -\mathcal{V}_{\nu\mu}^*$.
	Equation~\eqref{pertE}, valid in the weak-$\gamma$ regime away from EPs, shows that the level attraction between bands coupled through $V^{\rm (i)}(z)$ is determined by $|\mathcal{V}_{\mu\nu}|^2/|\Delta_{\mu\nu}|$.
	The pair that first reaches the EP is therefore the one with the largest coupling strength relative to the gap.
	For the lowest band, $\partial E_1/\partial\gamma>0$ implies a monotonic upward shift.
	For higher bands, the wave function overlap and the level spacing together determine the shift. Generally, the wave functions of the lowest few levels are concentrated near the minima of $V^{\rm (r)}(z)$; within each moir\'e unit cell, these minima satisfy $q\sin z=-\sin(z/q)$ and occur in $\mathcal{PT}$-conjugate pairs $z_n\Leftrightarrow z_{q-n+1}$, with $V(z_n)=V^*(z_{q-n+1})$.
	For even $q$, all minima form $q/2$ conjugate pairs.
	Taking $q=2$ as an example, $\tilde{\psi}_1$ and $\tilde{\psi}_2$ are mainly constructed from orbitals localized around the paired minima
	at $z_1/A\approx 0.3$ and $z_2/A \approx 0.7$, as shown in Fig.~\ref{fig3}(a1), forming a bonding-antibonding pair analogous to a $\mathcal{PT}$-symmetric double well \cite{HaagWunner2014PRA,DizdarevicWunner2015PRA}.
	The same concentration pattern persists in the non-Hermitian regime as seen in Fig.~\ref{fig3}(a2).
	Since $\tilde{\psi}_1$ and $\tilde{\psi}_2$ share the most similar concentration profiles, their energies are close ($\Delta_{21}<\Delta_{32}$) and $|\mathcal{V}_{21}|>|\mathcal{V}_{32}|$, hence $\partial E_1/\partial\gamma \approx -\partial E_2/\partial\gamma > 0$ at weak $\gamma$.
	The resulting strong attraction between the lowest two bands triggers $\mathcal{PT}$-symmetry breaking first at the moir\'e Brillouin boundary, where the gap is smallest; $D_2$ therefore increases monotonically with $\gamma$ in Fig.~\ref{fig2}(h).
	Odd $q$ has a different $\mathcal{PT}$-breaking sequence due to the self-conjugate center minimum $V^{\rm (i)}(A/2) = 0$. 
	For $q=3$, as shown in Fig. \ref{fig3}(b), $\widetilde\psi_2$ and $\widetilde\psi_3$ are concentrated on the same pair of off-center minima $z_1/A\approx0.2$ and $z_3/A\approx0.8$, producing a large $V^{\rm (i)}$-weighted off-diagonal matrix element $|\mathcal V_{32}|$. 
	By contrast, the center concentration of $\widetilde\psi_1$ suppresses $|\mathcal V_{21}|$, hence $\partial E_2/\partial\gamma \approx -\partial E_3/\partial\gamma > 0$, driving $\mathcal{PT}$-symmetry breaking between $E_{\nu=2}$ and $E_{\nu=3}$. 
	The same perturbative picture extends to general denominators. 
	For even $q_e$, the strongest coupling relative to the band gap occurs between the two lowest bands. Therefore, $\mathcal{PT}$-symmetry breaking first occurs in this band pair, and $D_{q_e}$ increases monotonically. 
	For odd $q_o$, the self-conjugate center minimum weakens the coupling of the lowest band and favors $\mathcal{PT}$-symmetry breaking in the higher off-center band pairs. 
	The tight-binding analysis in Appendix~\ref{AP:TB} shows that the self-conjugate central site couples only indirectly to the imaginary potential for odd $q_o$, whereas the lowest doublet couples directly for even $q_e$. This structural difference determines the first $\mathcal{PT}$-breaking band pair.
	
	On the other hand, for odd $q_o$, the nonmonotonic behaviors in Fig.~\ref{fig2}(g) arise from the competition between level attraction and non-Hermitian confinement. 
	At weak $\gamma$, attraction from higher bands broadens the lowest band, while $\mathcal{PT}$-symmetry breaking among these higher bands produces sharp $G_{q_o}$ features and subsequently weakens this attraction.
	As $\gamma$ grows further, the imaginary part of the potential dominates \cite{PelinovskyFrantzeskakis2013EL,Lin2025arxiv} and concentrates $\psi_1$ toward the center of each unit cell. This non-Hermitian confinement is evidenced by the decreasing wave-packet width $\sigma$ in Fig.~\ref{fig3}(c) and the increasingly concentrated wave function for $q=5$ in Fig.~\ref{fig3}(d), while its behavior at large $\gamma$ is analyzed further in Appendix~\ref{AP:hopeff}. It eventually overcomes the level-attraction-induced broadening and causes $D_5$ to decrease.
	
	\section{Interaction effect on the lowest moir\'e band}
	\begin{figure*}[th]	
		\includegraphics[width=1\linewidth]{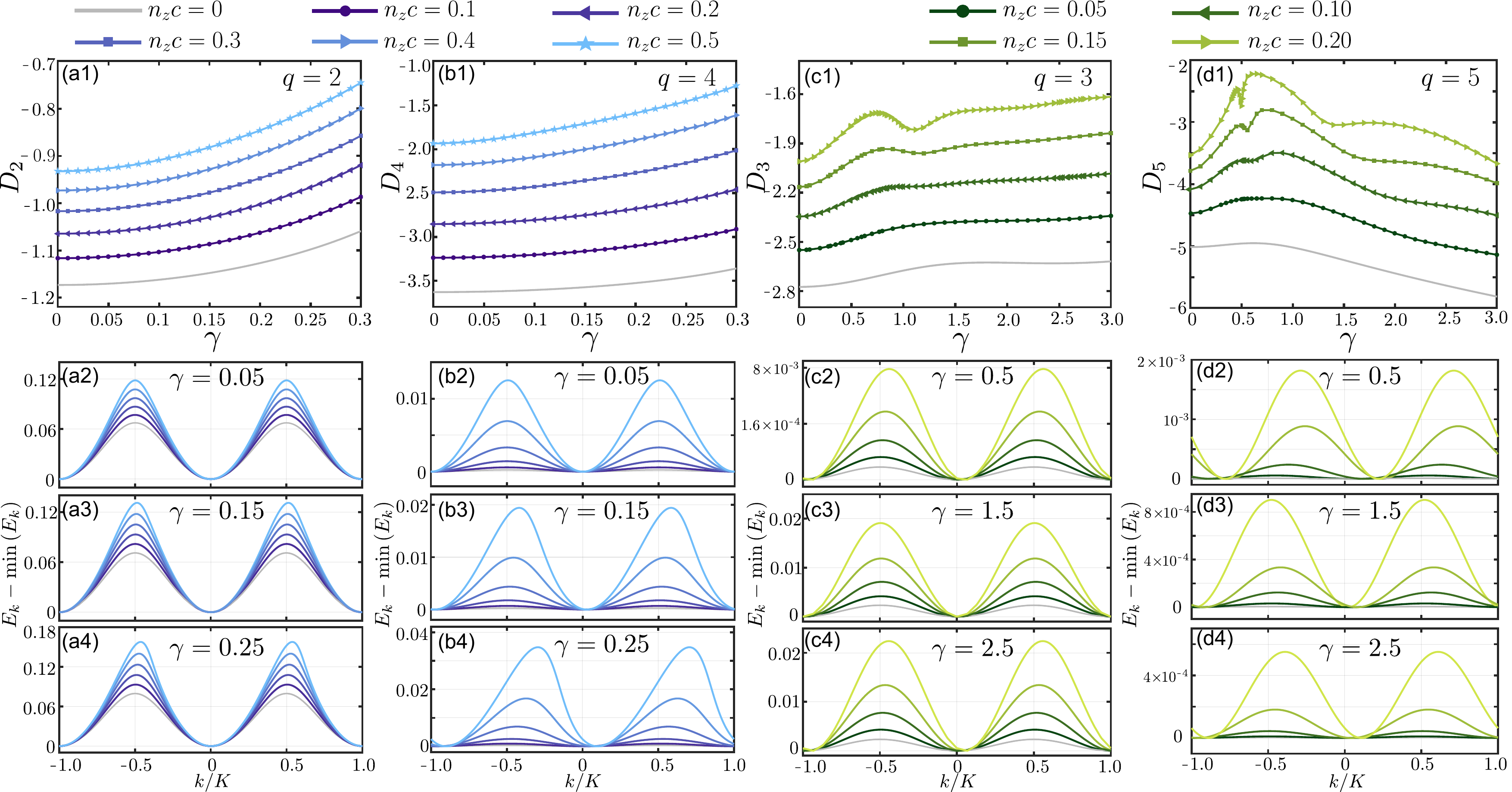}
		\caption{The lowest mean-field band and its dispersion $D_q$ as functions of the non-Hermitian strength $\gamma$, for different interaction strengths $n_z c$ (color coded). 
			Results for even parities ($q=2, 4$) are shown in (a)-(b), using a blue color scale for $n_z c = 0.1-0.5$. (a1),(b1) $D_{2,4}$ versus $\gamma$, while the typical band structures at $\gamma = \{0.05, 0.15, 0.25\}$ are displayed in (a2)-(a4) and (b2)-(b4). 
			Similarly, (c)-(d) present the odd parities ($q=3, 5$), with $n_z c = 0.05-0.2$ indicated by a green scale. The $D_{3,5}$ versus $\gamma$ relation is in (c1)(d1), and the band structures at $\gamma = \{0.5, 1.5, 2.5\}$ are in (c2)-(c4) and (d2)-(d4).}
		\label{fig4}
	\end{figure*}
	
	This section examines the effects of interaction strength $n_zc$ and non-Hermiticity $\gamma$ on the mean-field band structure and its dispersion within the GPE framework of a $\mathcal{PT}$ moir\'e lattice. 	
	We focus on weak repulsive interactions ($n_zc > 0$) to avoid the emergence of mean-field band swallowtails at the boundaries of the Brillouin zone. 
	For each $q$, we restrict $\gamma$ below the point at which the branch originating from the noninteracting lowest band either undergoes $\mathcal{PT}$-symmetry breaking for even $q_e$, or loses the lowest-real-part ordering through the crossing at $\gamma_c$ for odd $q_o$.
	For single-lattice systems, it is often possible to approximate $\psi(z)$ analytically using a superposition of a few plane waves for generic analysis \cite{WuNiu2003NJP, ZhangKonotop2021PRL}.
	However, the moir\'e lattice potential contains significantly more plane-wave components, making such trial wave functions inadequate. 
	Therefore, we numerically compute directly the self-consistent nonlinear solutions of Eq.  \eqref{timeIndepA}, setting the plane-wave momentum cutoff to $10qK$ to ensure accuracy (see Appendix \ref{APPNumerical} for details).
	
	Figure \ref{fig4} presents our numerical results, with the even ($q_e = \{2,4\}$) and odd ($q_o = \{3,5\}$) parities discussed separately. 
	For the even parities, shown in Figs. \ref{fig4}(a1)–\ref{fig4}(b1), the interaction-dependent dispersions $D_{q_e}$ (blue lines) exhibit a monotonically increasing trend with $\gamma$, consistent with the non-interacting case (gray lines). 
	As expected, the presence of interactions enhances band dispersion. 
	Interestingly, under the combined effect of interactions and non-Hermiticity, the band becomes asymmetric (i.e., $E_k \neq E_{-k}$), yielding an asymmetric band structure reminiscent of the single-lattice model discussed in Ref. \cite{ZhangKonotop2021PRL}.  
	To illustrate this clearly, Figs. \ref{fig4}(a2)–\ref{fig4}(a4) and Figs. \ref{fig4}(b2)–\ref{fig4}(b4) display the lowest real bands for $\gamma = \{0.05, 0.15, 0.25\}$ under different interaction strengths. 
	Both stronger interactions and larger non-Hermiticity enhance band asymmetry, which shifts the band maximum and ground state away from $k = K/2$ and $k = 0$, respectively, and also increase the band dispersion.
	Moreover, bands with a larger moir\'e unit cell ($q=4$) show greater sensitivity to interactions and non-Hermiticity than those with a smaller cell ($q=2$). 
	At the same $\gamma$, the bandwidth of $q=4$ [Fig. \ref{fig4}(b2)] increases significantly with interaction strength, in contrast to the modest increase for $q=2$ [Fig. \ref{fig4}(a2)]. 
	Similarly, for fixed $n_zc$, with increasing $\gamma$, band asymmetry is more pronounced for $q=4$ [Figs. \ref{fig4}(b2) and \ref{fig4}(b4)] than for $q=2$ [Figs. \ref{fig4}(a2) and \ref{fig4}(a4)], further reflecting the heightened sensitivity of the larger moir\'e unit cell to non-Hermitian effects.
	
	For odd parities, interactions enhance the dispersion while the nonmonotonic dependence of $D_{q_o}$ on $\gamma$ persists and develops local dips, most pronounced for $q=5$ at intermediate $\gamma$, as shown in Figs.~\ref{fig4}(c1) and \ref{fig4}(d1).
	The origin of this complex behavior can be traced through the interaction strength $n_z c$ as a control parameter. 
	At $n_zc = 0$, $D_{q_o}$ are governed by the $\mathcal{PT}$-breaking of $E_{\nu=2}$ and $E_{\nu=3}$ alone, inheriting the band-correlation rules in Sec. \ref{Sec.linear}. 
	For small $n_zc \lesssim 0.1$, the dispersion curves retain this structure, remaining smooth and undistorted [Figs.~\ref{fig4}(c2)--\ref{fig4}(c4)], where the noninteracting picture	remains qualitatively valid.
	For larger $n_zc$, the repulsive interaction shifts the nonlinear bands upward and reduces some interband gaps. 
	The enhanced influence of several higher-band branches amplifies the nonmonotonic behavior inherited from the noninteracting system, producing the pronounced oscillations and dips in Figs.~\ref{fig4}(c1) and \ref{fig4}(d1).		
	The lowest real bands for $\gamma = \{0.5, 1.5, 2.5\}$ are presented in Figs. \ref{fig4}(c2)–\ref{fig4}(c4) and \ref{fig4}(d2)–\ref{fig4}(d4), respectively, where band asymmetry induced by interactions and non-Hermiticity is again observed. 
	Unlike before, as $\gamma$ increases, the ground state exhibits oscillatory shifts toward $K$ and $-K$ at the corresponding $k$ points, a behavior generally accompanied by a noticeable enhancement in band dispersion. 
	
	\section{Conclusion and Discussion}\label{sec:conclusion}
	In summary, we investigated how non-Hermiticity and interactions modify the flatness of the lowest band in a $\mathcal{PT}$-symmetric moir\'e lattice. 
	In the noninteracting regime, the parity of the denominator $q$ in the commensurate ratio determines the first $\mathcal{PT}$-breaking pair and the resulting response of the lowest-band flatness. 
	For ratios with even denominators, the lowest two bands break $\mathcal{PT}$-symmetry first, and their level attraction produces monotonic broadening. 
	For ratios with odd denominators, the second and third lowest bands undergo $\mathcal{PT}$ symmetry breaking, while the lowest band remains real over a broad range of non-Hermiticity.
	The subsequent competition between level attraction and non-Hermitian confinement gives rise to a nonmonotonic response. 
	This parity-dependent phenomenon can be understood from perturbation theory. 
	Within the GPE framework, interactions generally enhance the mean-field band dispersion. For even denominators, interactions and non-Hermiticity both weaken the moir\'e-induced band flattening. 
	For odd denominators, however, non-Hermiticity can either enhance or suppress the flattening, leading to a richer interplay of cooperation and competition with interactions.
	
\begin{figure}[tb]	
		\includegraphics[width=1\linewidth]{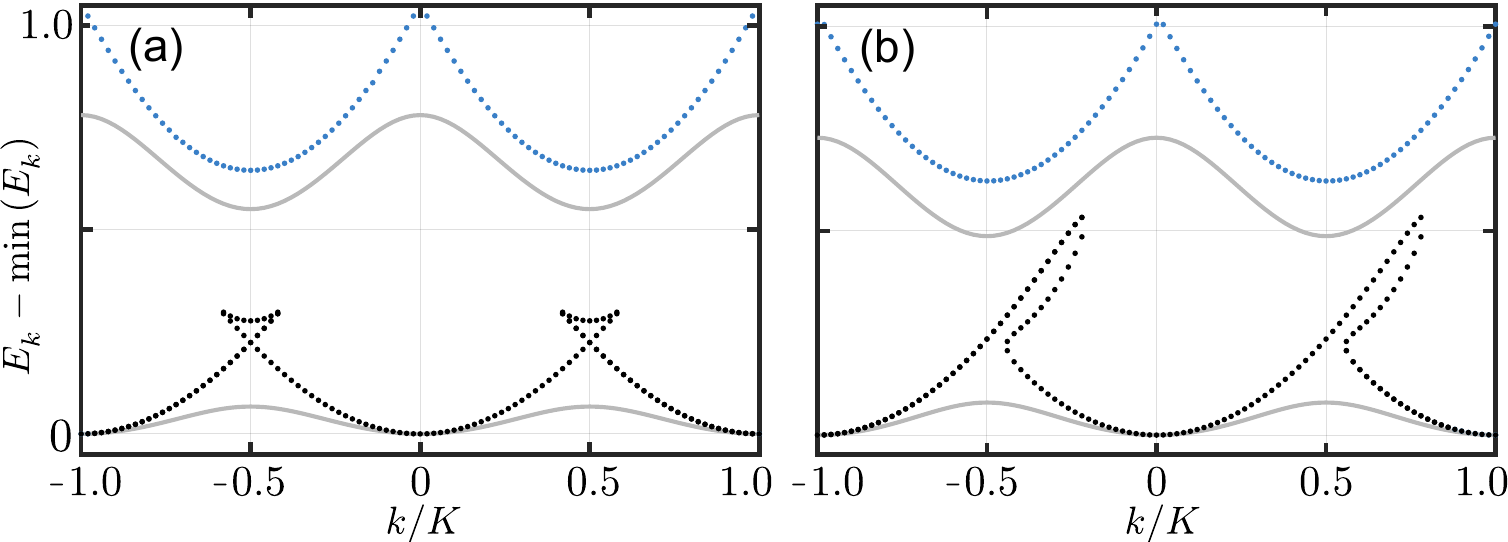}
		\caption{The lowest (black dots) and second (blue dots) mean-field bands at interaction strength $n_z c = 2$ for $q=2$ in (a) Hermitian and (b) non-Hermitian with $\gamma = 0.25$. 
			The linear bands are plotted in gray for comparison.}\label{figD}
	\end{figure}

	Although we only demonstrate the mean-field bands at weak interaction strengths, the strongly interacting regime featuring nonlinear swallowtail structures can also be numerically solved.	
	By iterating over a large number of random trial solutions, we obtained the swallowtail bands of the $\mathcal{PT}$-symmetric moir\'e lattice with $q=2$, for the Hermitian [Fig. \ref{figD}(a)] and non-Hermitian [Fig. \ref{figD}(b)] cases, respectively. 
	Specifically, non-Hermiticity appears to disrupt the swallowtail structure by removing one of the connecting points of the looped spectrum, in contrast to the single-lattice model \cite{ZhangKonotop2021PRL}.
	Given this peculiar multivalued band structure, it would be valuable to further investigate the corresponding dynamics, such as adiabatic or non-adiabatic Landau–Zener tunneling under a nonlinear sweep \cite{VitanovSuominen1999PRA,WangJie2022PRA}.
	Moreover, for other moir\'e ratios, how non-Hermiticity and interactions affect the existence of swallowtail structures is also an open and meaningful question to explore.
	
	On the other hand, the experimental realization of a BEC in such complex a potential  \eqref{OL} may become feasible in the future.
	A real bichromatic optical lattice can be generated by superimposing two standing-wave laser potentials with different periods \cite{Roati2008Roati}, and the effective 1D interaction strength can be tuned via Feshbach resonance by adjusting the $s$-wave scattering length \cite{TiesingaStoof1993PRA, InouyeKetterle1998nature}. 
	Moreover, imaginary periodic potentials implemented in coherently prepared $n$-level atomic vapors have been reported \cite{ShengXiao2013PRA,WuLaRocca2014PRL,ZhangXiao2016PRL}. 
	Although there is no direct evidence that such configurations can support a stable BEC, controlled atomic gain \cite{Tsuno2025Tsuno} and loss \cite{LiLuo2019NatComm,TakasuTakahashi2020,RenJo2022NaturePhys,WangZhao2024PRL} among hyperfine levels have been progressively achieved. 
	Along with the rapid advances of the ultracold atomic experimental techniques, the realization of $\mathcal{PT}$-symmetric lattices with periodic gain and loss can be expected experimentally.

	\begin{acknowledgments}
		We thank Kangwu Zheng and Cong-Jun Zou for valuable discussions.
		This work was supported by the National Key Research and Development Program of China (Grant No.~2022YFA1405304), the National Natural Science Foundation of China (Grant No.~12674331), the Guangdong Basic and Applied Basic Research Foundation (Grant No.~2024A1515010188), and the Startup Fund of South China Normal University.
	\end{acknowledgments}
	
	\section*{DATA AVAILABILITY}
	The data that support the findings of this article are openly available \cite{Cheng2026ChengData}.
	
	\appendix
	\section{Numerical methods for mean-field energy bands}\label{APPNumerical}
	The energy per particle $E$ can be obtained by first solving the nonlinear equation \eqref{timeIndepA}. We define the operator
	\begin{eqnarray}
		H[\psi_k] = \left[-4\left(\dfrac{\partial}{\partial z} + ik\right)^2 + V(z) + c|\psi_{k}|^2\right],
	\end{eqnarray}
	which satisfies $H[\psi_k] \psi_{k} = \mu_k \psi_{k}$. 
	Because of the periodicity of $\psi_{k}$, we expand it in a plane wave basis:
	\begin{eqnarray}
		\psi_{k}(z) = \sqrt{n_z} \sum_{l=-d}^{d} a_{k;l}e^{ilKz}, 
		\label{planWave}
	\end{eqnarray}
	where the integer $d$ is the cut off constant. 
	Substituting Eq. \eqref{planWave} into Eq. \eqref{normpsi} yields the normalization condition for the coefficients:
	\begin{eqnarray}
		\sum_{l=-d}^{d} |a_{k;l}|^2 = 1.
		\label{planWaveNorm}
	\end{eqnarray}
	In this plane-wave basis, we can express $H[\psi]$ for $\alpha = p/q$ as a matrix, whose elements are:
	\begin{eqnarray}
		\notag
		H_{k;n,m} &=& 4{\left(nK+k\right)^2}\delta_{n,m} + \frac{V_0}{2}\left(\delta_{n,m+q} + \delta_{n,m-q}\right)\qquad\\
		&\quad& + ~\frac{V_0}{2}(1+\gamma)\delta_{n,m+p} + \frac{V_0}{2}(1-\gamma)\delta_{n,m-p}\qquad\\\notag
		&\quad& + ~ n_zc \sum_{l^\prime,l} a_{k;l^\prime}^* a_{k;l} \delta_{l^\prime+n,l+m}.
	\end{eqnarray}
	Consequently, the time-independent GPE \eqref{timeIndepA} becomes a complex nonlinear system of equations
	\begin{eqnarray}
		\sum_{m = -d}^d H_{k;n,m}a_{\nu,k;m} = \mu_{\nu,k}a_{\nu,k;n},
		\label{matrixTimeIndep}
	\end{eqnarray}
	and the energy-per-particle functional \eqref{varepsilonA} for the interacting cases can be derived
	\begin{equation}
		E_k =	\mu_k - \dfrac{n_z c}{2}\sum_{n=-2d}^{2d} \left| \sum_{m = -d}^d a_{k,m} a_{k,n - m}\right|^2.
	\end{equation}
	
	In the non-interacting limit ($c=0$), Eq. \eqref{matrixTimeIndep} reduces to a non-Hermitian eigenvalue problem, which can be solved straightforwardly.  
	However, for the interacting case ($c\ne0$), we use the numerical iteration method to solve the complex nonlinear system of equations \eqref{matrixTimeIndep} under the normalization constraint \eqref{planWaveNorm}. 
	We decompose the $(2d+1)$ equations \eqref{matrixTimeIndep} into their real and imaginary parts since the coefficients $a$ are generally complex. 
	This yields a doubled system of equations over the real numbers
	\begin{eqnarray}
		\begin{bmatrix}
			H_k^{\rm (r)}-\mu_k&-H_k^{\rm (i)} \\
			H_k^{\rm (i)}&H_k^{\rm (r)}-\mu_k
		\end{bmatrix}
		\begin{bmatrix}
			\underline{a}^{\rm (r)}_k
			\\
			\underline{a}^{\rm (i)}_k
		\end{bmatrix}=0
		\label{matrixCompex}
	\end{eqnarray}
	where $\underline{a}_k$ is the vector of plane-wave coefficients. 
	We choose a specific gauge to ensure $a_{\nu,k;d}^{\rm (i)} = 0$. 
	This reduces the number of free variables to $4d+2$, which matches the number of equations of the real system \eqref{matrixCompex}; the normalization condition \eqref{planWaveNorm} is imposed by renormalizing the coefficient vector during the iteration.
	
	We solve for the set $\{\underline{a},\mu\}$ using the $\mathit{fsolve}$ function in {MATLAB}$^\copyright$. 
	The initial guess for the solution vector $\left ( \underline{a}^{\rm (r)},\underline{a}^{\rm (i)},\mu \right )^{\rm T}$ is constructed as follows: for the lowest band [Fig. \ref{fig4}], we used the result from the linear case $(c=0)$ as the trial solution to first iterate for the ground state (assumed at $k=0$). 
	After successful iteration, this solution was then used as the trial solution for $k+\Delta k$, and the process was repeated until reaching $k=K$. Subsequently, we iterated backwards from $k=K$ to $k=0$. If the results from the forward and backward iterations coincide completely near $k=K/2$, we conclude that no nonlinear swallowtail structure has emerged.
	For higher bands [Fig. \ref{figD}], we use random initial conditions.  
	The guess for the coefficient vector $\{\underline{a}\}$ is normalized on the $(4d+1)$-dimensional hypersphere to satisfy the normalization condition \eqref{planWaveNorm}. 
	The initial guesses for $\mu$ are set to the linear result multiplied by a random number between $0$ and $10$. 
	The moir\'e lattice introduces additional plane-wave couplings, necessitating an expansion in a large number of plane waves. The cut-off was set to $d=10q$ to ensure convergence, defined by the following criteria: the plane-wave coefficients for the lowest band at the Brillouin zone boundary satisfy $|a_{1,K/2;\pm d}| < 10^{-13}$, and the energy difference satisfies $|E_{k=0} - E_{k=\pm K}| < 10^{-8}$. 
	\\
	
	\begin{figure}[tb]	
		\includegraphics[width=1\linewidth]{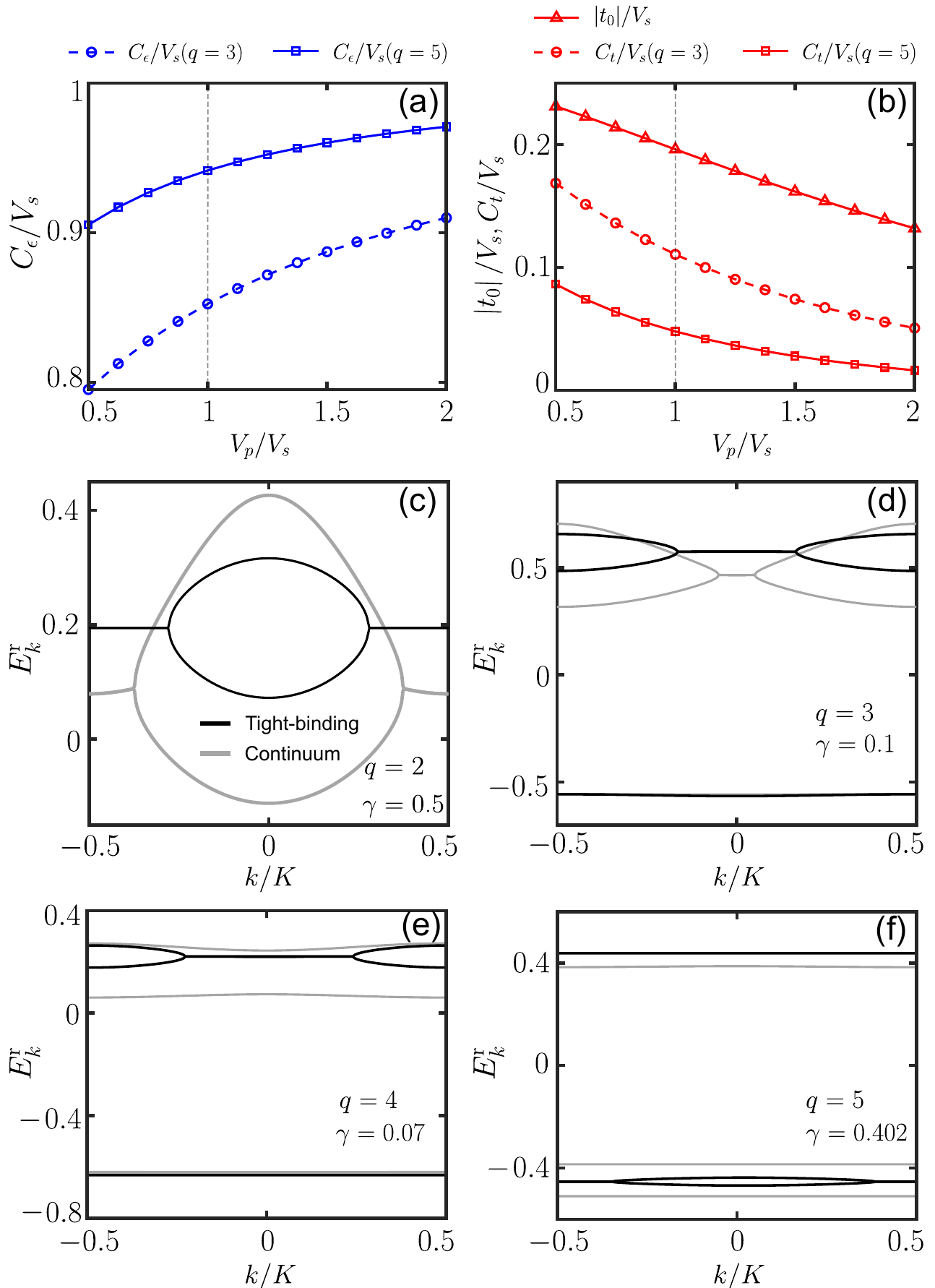}
		\caption{
			(a),(b) Tight-binding Hamiltonian parameters as functions of $V_p/V_s$ at fixed $V_s=0.8$, computed from the overlap integrals of the maximally localized Wannier functions via Eqs.~\eqref{TB_t0e0} and~\eqref{TB_C}. 
			(c)-(f)  Non-Hermitian band structures for $q=\{2,3,4,5\}$ at $V_p = V_s=0.8$.
			Tight-binding bands (black lines) are compared with the full continuum calculation performed using plane-wave-basis diagonalization (gray lines). 
			Panels (c) and (d) correspond to Class~{\rm I} ($\phi_p=0$, $\phi_s=0$);
			panels (e) and (f) correspond to Class~{\rm III} ($\phi_p=\pi$, $\phi_s=0$) as defined in Eq. \eqref{4cases_phi}.
		}\label{figAB}
	\end{figure}	
	
	\section{Tight-binding approximation for the continuum model}\label{AP:TB}
	In this Appendix, we discuss the non-interacting bands using a tight‑binding approach. 
	We rewrite the lattice potential in Eq. \eqref{lattice} as 
	\begin{eqnarray}
		\tilde{V}(z) = V_p \cos\left(z\right) + V_s \left[ \cos\left(\alpha z\right)
		+ i\gamma\sin\left(\alpha z\right) \right],\label{lattice_tbV}
	\end{eqnarray}
	and restrict ourselves to the case $V_p\ge V_s >0$. 
	Although the main text focuses on $\tilde{V}(z)$ with $V_p = V_s$, the key physical conclusions for weak $\gamma$ can be captured by Eq. \eqref{lattice_tbV}.
	In our setup the period of the primary lattice is always smaller than that of the secondary lattice.
	Consequently, the lowest band of the primary lattice undergoes significant folding once the secondary lattice is introduced.
	This folding occurs $q$ times, giving rise to $q$ moir\'e subbands in the lowest energy region, located in momentum space at $k_\nu = \pm\nu/2q$ ($\nu = 1, 2, \dots, q$). 
	Within the moir\'e Brillouin zone, gaps open between these subbands: 
	the gaps associated with odd $\nu$ are located at $k=\pm 1/2q$, whereas those associated with even $\nu$ appear at $k=0$. 
	Because the gaps among these $q$ lowest moir\'e subbands are much smaller than the gap above the lowest band of the primary lattice and can be estimated by perturbation theory, a tight-binding model based on the lowest band of the primary lattice provides an effective means of analyzing the moiré band structure.

	To construct the tight‑binding model, we identify the minima of the primary lattice (which nearly coincide with the moir\'e lattice minima \cite{GottlobSchneider2023PRB,JohnstoneSanchez2025PRA}) as the Wannier centers $R_s$. 
	The associated maximally localized Wannier functions centered at $R_s$ are denoted by $W(z-R_s)$ and can be chosen to be real and to have even parity \cite{Marzari2012Marzari}.
	Within one moir\'e unit cell, there are $q$ such centers, 
	\begin{eqnarray}
		R_s = (2s-1)\pi , \qquad s = 1,\dots,q. \label{Wcenter}
	\end{eqnarray}
	
	Let $(n,s)$ label the $s$ th site within the $n$ th moir\'e unit cell, and $c_{n,s}^{(\dagger)}$ be the corresponding annihilation (creation) operator. 
	The noninteracting  real‑space Hamiltonian with nearest-neighbor hopping takes the form
	\begin{equation}
		\label{TB_realH}
		H = \sum_{n,s} \tilde{\varepsilon}_s \, c_{n,s}^\dagger c_{n,s}
		+ \sum_{n,s} \bigl( \tilde{t}_s \, c_{n,s+1}^\dagger c_{n,s} + \text{h.c.} \bigr),
	\end{equation}
	where the onsite energies and nearest-neighbor hopping amplitudes are defined as
	\begin{equation}
		\begin{aligned}
			\tilde{\varepsilon}_s &\equiv  \int_{-\infty}^{+\infty}W(z-R_s)\left[-4\frac{\partial^2}{\partial z^2} + \tilde{V}(z)\right]W(z-R_s)\,dz,\\ 
			\tilde{t}_s &\equiv \int_{-\infty}^{+\infty}W(z-R_{s+1})\left[-4\frac{\partial^2}{\partial z^2} +\tilde{V}(z) \right]W(z-R_s)\,dz.
		\end{aligned}\label{onsite_hopping}
	\end{equation}
	Taking into account the parity of the integrands, these expressions can be reduced to
	\begin{equation}
		\begin{aligned}
			\tilde{\varepsilon}_s &= \varepsilon_0 +  C_\varepsilon\Bigl[\cos\Bigl(\frac{R_s}{q}\Bigr) + i\gamma\sin\Bigl(\frac{R_s}{q}\Bigr)\Bigr], \\
			\tilde{t}_s &= t_0 + C_t\Bigl[\cos\Bigl(\frac{R_s+\pi}{q}\Bigr) + i\gamma\sin\Bigl(\frac{R_s+\pi}{q}\Bigr)\Bigr],
		\end{aligned}
		\label{TB_st}
	\end{equation}
	where $\varepsilon_0, t_0 \in \mathbb{R}$ are the onsite energy and nearest-neighbor hopping amplitude of the primary lattice, respectively, and are given explicitly by
	\begin{equation}
		\begin{aligned}
			\varepsilon_0 &= \int_{-\infty}^{+\infty}W(z-R_s)\left[-4\frac{\partial^2}{\partial z^2} + V_p \cos z\right]W(z-R_s)dz, \\
			t_0 &= \int_{-\infty}^{+\infty}W(z-R_{s+1})\left[-4\frac{\partial^2}{\partial z^2} + V_p \cos z\right]W(z-R_s)dz.
		\end{aligned}\label{TB_t0e0}
	\end{equation}
	The constants
	\begin{equation}
		\begin{aligned}
			C_\varepsilon &\equiv V_s\int_{-\infty}^{+\infty} W^2(u)\cos(u/q)du, \\
			C_t &\equiv V_s\int_{-\infty}^{+\infty} W(v+\pi)W(v-\pi)\cos(v/q)dv\label{TB_C}
		\end{aligned}
	\end{equation}
	are real numbers with $C_\varepsilon/V_s>0$ and $C_t/V_s>0$, whose precise values do not affect the symmetry content of Eq. \eqref{TB_st}. 
	Throughout this work, we restrict our discussion to 
	\begin{eqnarray}
		C_t < |t_0|, ~~~ t_0 < 0, \label{tce_con}
	\end{eqnarray}
	which ensures that the primary lattice tight-binding description underpinning Eq.~\eqref{TB_realH} remains valid. 
	The numerically computed tight-binding parameters across a range of primary and secondary lattice depths are shown in Figs. \ref{figAB} (a) and \ref{figAB} (b).
	
	One readily verifies the following relations
	\begin{eqnarray}
		\tilde{\varepsilon}_s = \tilde{\varepsilon}_{q-s+1}^*, \qquad \tilde{t}_s = \tilde{t}^*_{q-s},\label{pairingts}
	\end{eqnarray}
	which reveal a $\mathcal{PT}$-type pairing between sublattice $s$ and $q-s+1$; in particular, for odd $q$ one has $\tilde{\varepsilon}^{\rm (i)}_{(q+1)/2} = 0$.
	Moreover, denoting $\varphi_s \equiv \arg[\tilde{t}_s]$, the pairing relation in Eq. \eqref{pairingts} implies pairwise cancellation; hence, $\sum_{s=1}^q \varphi_s \equiv 0$ (mod $\pi$).
	Applying the Peierls transformation $c_{n,s} \to e^{i\theta_s} c_{n,s}$ with $\theta_{s+1} - \theta_s =\varphi_s$ (mod $\pi$) renders every hopping real \cite{Peierls1933Peierls,AshidaUeda2020}.
	After the transformation, $t_s = t_{q-s}$, while the onsite terms are unaffected: $\varepsilon_s = \tilde{\varepsilon}_s$.
	
	Applying a Fourier transform to Eq. \eqref{TB_realH}, the Bloch Hamiltonian in $k$ space reads
	\begin{align}
		H(k) =
		\begin{bmatrix}
			\varepsilon_1 & t_1 & 0 & \cdots & 0 & t_q e^{-i\theta_k} \\
			t_1 & \varepsilon_2 & t_2 & \cdots & 0 & 0 \\
			\vdots & \vdots & \vdots & \ddots & \vdots & \vdots \\
			0 & 0 & 0 & \cdots & \varepsilon_2^* & t_1 \\
			t_q e^{i\theta_k} & 0 & 0 & \cdots & t_1 & \varepsilon_1^*
		\end{bmatrix}_{q\times q},\label{HK}
	\end{align}
	where $\theta_k = 2\pi q k$. 
	Diagonalizing $H(k)$ at each $\gamma$ yields the moir\'e tight-binding band structure $\mathcal{E}_q$, from which $D_q$ and $G_q$ are extracted.
	
	\subsection{The even case: $q=2$}
	For $q=2$ the tight-binding chain reduces to the non-Hermitian Su-Schrieffer-Heeger model \cite{LangChong2018PRB,AshidaUeda2020}, whose eigenvalues can be obtained analytically:
	\begin{equation}
		E_\pm(k) = \varepsilon_0 \pm \sqrt{2(t_0^2+C_t^2)+2\left(t_{0}^{2}- C_{t}^{2}\right) \cos \theta_k-\gamma^{2} C_{\varepsilon}^{2}}.
	\end{equation}
	Figure \ref{figAB}(c) compares the tight-binding bands (black) with the continuum bands (gray) at $\gamma=0.5$.
	
	Under the condition in Eq. \eqref{tce_con}, the $\mathcal{PT}$-symmetry breaking threshold is
	\begin{equation}
		\gamma_{pt}(k) = \dfrac{\sqrt{2(t_0^2+C_t^2)+2\left(t_{0}^{2}- C_{t}^{2}\right)\cos\theta_k }}{C_\varepsilon}.
	\end{equation}
	The earliest $\mathcal{PT}$-symmetry breaking occurs at the moir\'e Brillouin-zone boundary, $k=\pm1/4$, where $\gamma_{pt}(\pm1/4) = 2C_t/C_\varepsilon$.   	
	Numerical integration with the maximally localized Wannier functions yields $\gamma_{pt}(1/4)\approx 0.48$ for $V_p=V_s=0.8$, slightly above the continuum-model value $0.46$ shown in Fig. \ref{fig2}(d). 
	
	\subsection{The odd case: $q=3$}
	For $q=3$, the model is usually referred to as the $\mathcal{PT}$-trimer model  \cite{HangVladimir2013PRL,Jin2017PRA,GarmonNoba2021PRA,AnastasiadisFotios2022PRB,Yin2024arxiv}, where $\mathcal{PT}$ symmetry pairs sublattices $|s=1\rangle \leftrightarrow |3\rangle$, while $|2\rangle$ is self-conjugate. 
	Using the pairing unitary matrix 
	\begin{eqnarray}
		U = \frac{1}{\sqrt{2}}\begin{bmatrix}
			1& 0 &-1\\
			1& 0 &1 \\
			0& \sqrt{2} &0\end{bmatrix},\quad
	\end{eqnarray}
	which satisfies $(|-\rangle, |+\rangle, |2\rangle)^{\rm T} = U(|1\rangle, |2\rangle, |3\rangle)^{\rm T}$ with $|\pm\rangle \equiv (|1\rangle \pm |3\rangle)/\sqrt{2}$, the transformed Hamiltonian $\mathcal{H}(k) = U H(k) U^{\rm T}$ reads
	\begin{eqnarray}
		\mathcal{H}(k) = 
		\begin{bmatrix}
			\varepsilon_1^{\rm (r)}-t_3\cos\theta_k& i(\varepsilon _1^{\rm (i)}-t_3\sin\theta_k) &0\\
			i(\varepsilon _1^{\rm (i)}+t_3\sin\theta_k) & \varepsilon_1^{\rm (r)}+t_3\cos\theta_k & \sqrt{2}t_1 \\
			0& 	\sqrt{2}t_1&\varepsilon_2 
		\end{bmatrix},\qquad~~\label{mHkq3}
	\end{eqnarray}
	where $\varepsilon_1 = \varepsilon_0+C_\varepsilon(1+i\sqrt{3}\gamma)/2$ and $\varepsilon_2 = \varepsilon_0-C_\varepsilon$.
	
	We first consider the Hermitian limit $\gamma = 0$. 
	At the moir\'e zone center $k = 0$, the matrix becomes block diagonal: $|-\rangle$ decouples completely, with eigenvalue $E_{|-\rangle} = \varepsilon_1^{\rm{(r)}} - t_3$, while $|+\rangle$ and $|2\rangle$ form a coupled $2\times 2$ subsystem whose eigenvalues we denote by $E_\pm = (\varepsilon_1^{\rm (r)} + \varepsilon_2  + t_3 \pm D)/2$ where $D \equiv \sqrt{(3C_\varepsilon/2 + t_3)^2 + 8t_1^2} > 0$. 
	
	Under the condition in Eq.~\eqref{tce_con}, one has $t_3 < 0$ in the weak-$\gamma$ tight-binding regime. 
	Together with $C_\varepsilon \gg C_t$ (the onsite term induced by the
	secondary lattice dominates over the hopping modulation [Figs.~\ref{figAB}(a) and (b)]), a direct calculation yields
	\begin{eqnarray}
		E_{|-\rangle} > E_+ > E_-,\qquad  \Delta_{21} > \Delta_{32}.
	\end{eqnarray}
	where $\Delta_{21} \equiv E_+ - E_-$ and $\Delta_{32} \equiv E_{|-\rangle} - E_+$ follow the band-index convention of Sec. \ref{Sec.linear}. 
	The two upper bands (the decoupled $|-\rangle$ and the anti-bonding combination of $|+\rangle$ and $|2\rangle$) are split by the smaller gap $\Delta_{32}$, while the lowest band, dominated by the self-conjugate center sublattice $|2\rangle$, is separated from them by the larger gap $\Delta_{21}$. 
	This inverted gap hierarchy is the key structural feature that governs the $\mathcal{PT}$-breaking sequence. 
	
	When $\gamma \neq 0$, the imaginary onsite contrast generates the coupling $\varepsilon_1^{\rm (i)} = \gamma C_\varepsilon\sqrt{3}/2$ between $|+\rangle$ and $|-\rangle$.
	To analyze the $\gamma$ dependence of the lowest level, we take $k=0$ and consider the characteristic equation $\det[\mathcal{H} - E I] = 0$. 
	A direct expansion yields
	\begin{equation}
		(E - \varepsilon_2)\bigl[(E - \varepsilon_1^{\rm (r)})^2 - t_3^2 + (\varepsilon_1^{\rm (i)})^2\bigr] = 2t_1^2\,(E - \varepsilon_1^{\rm (r)} + t_3).
		\label{detEq3}
	\end{equation}
	The lowest level is close to $\varepsilon_2$. 
	We define $E_- = \varepsilon_2 +\delta E$ and $|\delta E| \ll \Delta_\varepsilon$, where $\Delta_\varepsilon \equiv \varepsilon_1^{\rm (r)} - \varepsilon_2$.  
	In the tight-binding regime, Eq.~\eqref{detEq3} yields, to leading order in $t_1^2$,
	\begin{eqnarray}
		\delta E \approx -\dfrac{2t_1^2(\Delta_\varepsilon - t_3)}{\Delta_\varepsilon^2 - t_3^2 + (\varepsilon_1^{\rm (i)})^2}.
	\end{eqnarray}
	The numerator contains a $\gamma$-dependent term that scales as $\gamma^2 C_t^2$, while the denominator contains $(\varepsilon_1^{\rm (i)})^2 \propto \gamma^2 C_\varepsilon^2$.
	Since $C_\varepsilon^2 \gg C_t^2$ [Figs.~\ref{figAB}(a) and (b)], the denominator grows much faster with $\gamma$ than the numerator does;
	consequently $|\delta E|$ decreases and the real part of $E_-$ shifts upward, consistent with the continuum model behavior shown in Fig.~\ref{fig2}(b).
	This suppression of the level repulsion signals a non-Hermitian-induced level attraction in the weak-$\gamma$ region. 
	
	For the two upper levels, the $|\pm\rangle$ sector determinant in Eq. \eqref{mHkq3} controls the $\mathcal{PT}$-symmetry breaking. 
	The two eigenvalues $E_{|-\rangle}$ and $E_+$ satisfy $(E - \varepsilon_1^{\rm (r)})^2 \approx t_3^2 - (\varepsilon_1^{\rm (i)})^2$ (neglecting the coupling $t_1$), and coalesce when $\varepsilon_1^{\rm (i)} = |t_3|$, i.e., $\gamma_{pt} \approx 2|t_3|/(C_\varepsilon\sqrt{3})$. 
	
	\subsection{The general case}
	The $\mathcal{PT}$ pairing $|s\rangle\leftrightarrow|q-s+1\rangle$ encoded in Eq.~\eqref{TB_st} has fundamentally different consequences for odd and even $q$. For odd $q=2m+1$, the sublattices in each moir\'e unit cell form $m$ $\mathcal{PT}$-conjugate pairs together with one additional self-conjugate center sublattice $s_c=(q+1)/2=m+1$ at $R_{s_c}=q\pi$. 
	From Eq.~\eqref{TB_st}, its onsite energy satisfies $\varepsilon_{s_c}=\varepsilon_0-C_\varepsilon$ and $\varepsilon_{s_c}^{\rm (i)}=0$, and is therefore purely real. 
	For even $q$, no self-conjugate center sublattice exists, and every sublattice belongs to a $\mathcal{PT}$-conjugate pair.
	
	To determine the coupling of the center sublattice for odd $q$, we introduce the pairing basis $|s,\pm\rangle\equiv(|s\rangle\pm|q-s+1\rangle)/\sqrt{2}$, with $s=1,\ldots,m$. 
	In the original site basis, the center state $|s_c\rangle=|m+1\rangle$ couples only to its nearest neighbors $|m\rangle$ and $|m+2\rangle$. 
	These two states form the last conjugate pair $|m,\pm\rangle=(|m\rangle\pm|m+2\rangle)/\sqrt{2}$. 
	Because the two corresponding hoppings are equal under the $\mathcal{PT}$ pairing, their antisymmetric contributions cancel, and the center state couples only to $|m,+\rangle$. Equivalently, $\langle m,+|H|s_c\rangle=\sqrt{2}t_m$, whereas $\langle m,-|H|s_c\rangle=0$.
	
	In the paired basis $\left(|1,-\rangle,|1,+\rangle,\ldots,|m,-\rangle,|m,+\rangle,|s_c\rangle\right)^{\rm T}$, the odd-$q$ Bloch Hamiltonian takes the block-tridiagonal form
	\begin{eqnarray}
		\mathcal{H}(k)=
		\begin{bmatrix}
			\mathbf{H}_1(k) & \mathbf{T}_{12} & \mathbf{0}
			& \cdots & \mathbf{0} \\
			\mathbf{T}_{12}^{\rm T} & \mathbf{H}_2 & \ddots
			& & \vdots \\
			\mathbf{0} & \ddots & \ddots & \ddots
			& \mathbf{0} \\
			\vdots & & \ddots & \mathbf{H}_m
			& \mathbf{V}_{m,c} \\
			\mathbf{0} & \cdots & \mathbf{0}
			& \mathbf{V}_{m,c}^{\rm T} & \varepsilon_{s_c}
		\end{bmatrix}.
	\end{eqnarray}
	Here, each boldface diagonal block $\mathbf{H}_s$ is a $2\times2$ $\mathcal{PT}$-symmetric matrix associated with one conjugate sublattice pair. The only $k$-dependent block is $\mathbf{H}_1(k)$, which originates from the intercell hopping $t_q$ connecting the two sites of the outermost conjugate pair:
	\begin{eqnarray}
		\mathbf{H}_1(k)=
		\begin{bmatrix}
			\varepsilon_1^{\rm (r)}-t_q\cos\theta_k
			&
			i(\varepsilon_1^{\rm (i)}-t_q\sin\theta_k)
			\\
			i(\varepsilon_1^{\rm (i)}+t_q\sin\theta_k)
			&
			\varepsilon_1^{\rm (r)}+t_q\cos\theta_k
		\end{bmatrix}.
		\label{tbH1k}
	\end{eqnarray}
	The remaining diagonal blocks are $k$-independent and take the form
	\begin{eqnarray}
		\mathbf{H}_s=
		\begin{bmatrix}
			\varepsilon_s^{\rm (r)} & i\varepsilon_s^{\rm (i)}\\
			i\varepsilon_s^{\rm (i)} & \varepsilon_s^{\rm (r)}
		\end{bmatrix},
		\qquad s=2,\ldots,m.
		\label{tbHsk}
	\end{eqnarray}
	The inter-pair couplings and the coupling to the center site are $\mathbf{T}_{s,s+1}=t_s\mathbf{I}_2$ for $s=1,\ldots,m-1$ and $\mathbf{V}_{m,c}=\sqrt{2}t_m(0,1)^{\rm T}$, respectively.
	
	The form of $\mathbf{V}_{m,c}$ explicitly shows that the additional self-conjugate state $|s_c\rangle$ couples directly only to the symmetric state $|m,+\rangle$. Its coupling to the antisymmetric state $|m,-\rangle$ is indirect and occurs through the imaginary matrix element $i\varepsilon_m^{\rm (i)}$ in $\mathbf{H}_m$ [Eq.~\eqref{tbHsk}]. 
	Since $\varepsilon_{s_c}^{\rm (i)}=0$ and the lowest eigenstate is dominated by the self-conjugate center site in the lattice considered here, its first-order energy correction from the imaginary potential vanishes. 
	Its leading correction at weak $\gamma$ is therefore second order and real. Consequently, the lowest band remains real over a broader range of non-Hermiticity, while the directly coupled second and third bands reach $\mathcal{PT}$-breaking exceptional points first.
	
	For even $q$, the additional self-conjugate center site and the scalar block $\varepsilon_{s_c}$ are absent. The paired basis then consists entirely of $|s,\pm\rangle$ doublets, and the Hamiltonian contains only coupled $2\times2$ $\mathcal{PT}$-symmetric blocks. The lowest doublet therefore couples directly through the imaginary onsite terms and breaks $\mathcal{PT}$ symmetry first. Thus, the presence or absence of the additional self-conjugate center site determines the first symmetry-breaking band pair and provides the structural origin of the parity-dependent response of the lowest-band flatness.
	
	\subsection{Nontrivial relative phase in a moir\'e lattice}
	In experimental realizations of bichromatic optical lattices, the primary and secondary lattices generally possess independent relative phases with respect to the laboratory frame.
	The real part of the potential is accordingly generalized to 
	\begin{equation}
		\tilde{V}^{\rm (r)}(z) = V_p \cos\left(z+\phi_p\right) + V_s  \cos\left(\alpha z+\phi_s\right),\label{lattice_tbV2}
	\end{equation}
	with $V_p, V_s >0$.	
	The imaginary part, on the other hand, carries no phase degree of freedom; its form is determined by the effective non‑Hermitian control in experiments \cite{LiLuo2019NatComm,TakasuTakahashi2020,RenJo2022NaturePhys,WangZhao2024PRL,Tsuno2025Tsuno}
	\begin{eqnarray}
		\tilde{V}^{\rm (i)}(z) = \gamma V_I \sin\left(\alpha z\right), ~~V_I \equiv V_s.
	\end{eqnarray}
	
	Imposing $\mathcal{PT}$ symmetry, $\tilde{V}^*(-z)=\tilde{V}(z)$, constrains  $\phi_p=n_p\pi$ and $\phi_s=n_s\pi$ with $n_p,n_s\in\mathbb{Z}$.
	Modulo $2\pi$, this yields four distinct $\mathcal{PT}$-symmetric configurations,
	\begin{equation}\label{4cases_phi}
		(\phi_p,\phi_s)\in\left\{(0,0),~(0,\pi),~(\pi,0),~(\pi,\pi)\right\}.
	\end{equation}
	These four classes are inequivalent under spatial translations of the coordinate, and they correspond to the four distinct combinations of red‑ and blue‑detuned primary and secondary lattices:  
	\begin{itemize}[label={}, leftmargin=*]
		\item \label{4cases} I~$(0,0)$: ~~both blue-detuned;
		\item II~$(0,\pi)$: ~primary blue, secondary red-detuned;
		\item III~$(\pi,0)$: primary red, secondary blue-detuned;
		\item IV~$(\pi,\pi)$: both red-detuned.
	\end{itemize}
	
	We now use the tight-binding model to examine how the relative phases modify the low-energy $\mathcal{PT}$-breaking sequence in all four classes.
	
	We construct a single-band tight-binding model using the maximally localized Wannier functions of the primary lattice.
	The Wannier centers are determined by the minima of the primary potential:
	\begin{equation}
		R_s=
		\begin{cases}
			(2s-1)\pi, & ({\rm I,II}),\\
			2(s-1)\pi, & ({\rm III,IV}),
		\end{cases}
		\qquad s=1,\ldots,q.
		\label{fourcasecenters}
	\end{equation}
	The shape and parity of the Wannier functions are independent of $\phi_p$ and $\phi_s$ because the primary-lattice depth $|V_p|$ is unchanged.
	
	The tight-binding parameters for the four configurations are obtained from the overlap integrals defined in Eqs.~\eqref{onsite_hopping}--\eqref{TB_t0e0}.
	The primary-lattice phase $\phi_p$ determines the positions and $\mathcal{PT}$ pairing of the Wannier centers through Eq.~\eqref{fourcasecenters}, whereas the secondary-lattice phase $\phi_s$ changes the ordering of their real onsite energies.
	The imaginary potential itself remains unchanged because it is generated independently of the detuning configuration.
	
	The $\mathcal{PT}$-pairing relations are determined by $\phi_p$:
	\begin{equation}
		\begin{aligned}
			\phi_p=0~({\rm I,II}):\quad&
			\widetilde{\varepsilon}_s=\widetilde{\varepsilon}_{q-s+1}^{*},
			\quad\widetilde{t}_s=\widetilde{t}_{q-s}^{*};\\
			\phi_p=\pi~({\rm III,IV}):\quad&
			\widetilde{\varepsilon}_s=\widetilde{\varepsilon}_{q-s+2}^{*},\quad
			\widetilde{t}_s=\widetilde{t}_{q-s+1}^{*}.
		\end{aligned}
		\label{4casePT_pairings}
	\end{equation}
	The corresponding self-conjugate sites are
	\begin{equation}
		\begin{aligned}
			\phi_p = 0\;(\mathrm{I,II}):\quad
			&\tilde{\varepsilon}_{(q+1)/2} \in \mathbb{R},
			~~~(\mathrm{odd}\;q),&&\\
			\phi_p = \pi\;(\mathrm{III,IV}):\quad
			&\begin{cases}
				\tilde{\varepsilon}_{1} \in \mathbb{R},
				& (\mathrm{all}\;q),\\
				\tilde{\varepsilon}_{q/2+1} \in \mathbb{R},
				& (\mathrm{even}\;q).
			\end{cases}
		\end{aligned}
		\label{4casePT_self_conj}
	\end{equation}
	
	However, the existence of a self-conjugate site alone does not determine the low-energy $\mathcal{PT}$-breaking sequence.
	Such a site delays the breaking of the lowest band only when it belongs to the lowest-energy sector.
	The breaking pattern is therefore jointly controlled by the $\mathcal{PT}$-pairing geometry and the ordering of the real onsite energies.
	
	In Class I, the self-conjugate center site belongs to the lowest-energy sector for odd $q$, whereas no self-conjugate site exists for even $q$.
	Consequently, the second and third bands undergo $\mathcal{PT}$-symmetry breaking first for odd $q$, while the lowest two bands break first for even $q$, as discussed in the main text.
	
	In Class II, the phase shift $\phi_s=\pi$ reverses the ordering of the real onsite energies without changing the imaginary potential.
	For odd $q$, although a geometrically self-conjugate site still exists, it no longer corresponds to a lowest-energy minimum.
	For even $q$, no self-conjugate site exists.
	Therefore, for both odd and even $q$, the lowest states are mainly formed from $\mathcal{PT}$-conjugate sites, and the lowest two bands undergo $\mathcal{PT}$-symmetry breaking first.
	
	Class III provides the clearest contrast with Class I.
	The shift $\phi_p=\pi$ moves the primary-lattice Wannier centers by half a primary-lattice period relative to the unchanged imaginary potential.
	In this case, a lowest-energy self-conjugate site occurs for even $q$ but not for odd $q$, thereby reversing the parity assignment of Class I.
	Figures~\ref{figAB}(e) and \ref{figAB}(f) show representative tight-binding and continuum spectra for $q=4$ and $q=5$, respectively.
	For $q=4$ [Fig.~\ref{figAB}(e)], $\mathcal{PT}$-symmetry breaking first occurs between $E_2$ and $E_3$, while the lowest band remains real.
	For $q=5$ [Fig.~\ref{figAB}(f)], the lowest two bands form the first $\mathcal{PT}$-breaking pair.
	
	In Class IV, the secondary-lattice phase places at least one self-conjugate site in the lowest-energy sector for both odd and even $q$.
	The lowest band is therefore protected from the first $\mathcal{PT}$-breaking transition for both parities, and the first transition instead involves higher bands.
	
	The four configurations thus exhibit distinct low-energy breaking patterns.
	Class I shows the parity dependence discussed in the main text, while Class III reverses the roles of odd and even $q$.
	Classes II and IV remove the odd--even distinction: the lowest two bands break first for both parities in Class II, whereas the lowest band remains real at the first $\mathcal{PT}$-symmetry breaking for both parities in Class IV.
	Based on the connection established in Sec.~\ref{Sec.linear} between the first $\mathcal{PT}$-symmetry breaking pair and the lowest-band dispersion, we expect the associated monotonic and nonmonotonic responses to follow the same classification.
	For conciseness, only the representative Class III spectra are shown in Fig.~\ref{figAB}; the results for Classes II and IV are summarized above.
	
	\begin{figure}[tb]	
		\includegraphics[width=0.7\linewidth]{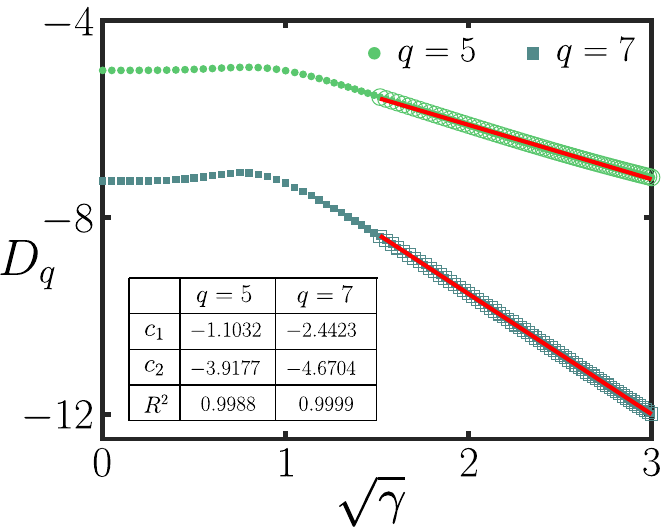}
		\caption{
			Over the fitting interval, $D_q$ is approximately linear in $\sqrt{\gamma}$ for $q=5$ and $q=7$ in the continuum model at $V_0=0.8$.
			The red lines are linear fits $D_q\simeq c_1\sqrt{\gamma}+c_2$ over the range $2\leq\gamma\leq9$, where we track the real branch continuously connected to the Hermitian lowest band. 
			This branch remains sufficiently separated from the other bands in the complex-energy plane.
			The lower‑left table lists the fitted parameters and the corresponding $R^2$ values for each $q$.
		}\label{figAC}
	\end{figure}

	\section{The effective hopping in the large-$\gamma$ regime}\label{AP:hopeff}
	The tight-binding analysis in Appendix \ref{AP:TB} relies on the lowest band of the primary lattice, for which only the lowest Wannier orbital of that lattice is retained; the secondary lattice enters solely through the matrix elements $C_\varepsilon$ and $C_t$.
	At large $\gamma$, this truncation becomes inadequate because the strength of the imaginary potential $i\gamma\sin(z/q)$ becomes comparable to the primary lattice band gap, thereby mixing higher Wannier orbitals into the lowest band states and suppressing propagation beyond what fixed hopping integrals can capture. 
	A clear manifestation of this breakdown is the monotonic decrease of $D_5$ with increasing $\gamma$ in Fig. \ref{fig2}(g), which signals the progressive failure of Eq. \eqref{TB_realH}.
	
	This section explains the flattening of the lowest band with increasing $\gamma$ for odd $q$.
	We construct a tight-binding model using Wannier functions obtained from the eigenstates of the full non-Hermitian Hamiltonian $\tilde{H} = -4\partial_z^2+\tilde{V}(z)$.
	The width of the lowest band $w_q\approx 4|t_{\rm eff}|$ is set by the effective hopping between equivalent Wannier centers in neighboring cells \cite{Gong2018Gong,AshidaUeda2020}. 
	The hopping integral is
	\begin{equation}
		t_{\rm eff}(\gamma) \equiv \int_{-\infty}^{\infty}[W_L(z-R_c)]^* \tilde{H} W_R(z+R_c) dz.\label{hopping_eff}
	\end{equation}
	Here, $W_L$ and $W_R$ denote the left and right Wannier functions constructed from the eigenstates of $\tilde{H}^{\dagger}$ and $\tilde{H}$, respectively.
	
	At large $\gamma$, the imaginary part of the potential dominates in the tail region away from the zeros of $\sin(z/q)$, so that $\tilde{V}(z)\approx i\gamma V_s\sin(z/q)$.
	Although this approximation is not locally valid near the zeros at $z=n\pi q$, including those corresponding to the Wannier centers and the midpoint between neighboring centers, these narrow matching regions contribute only subleading corrections to the large-$\gamma$ action. 
	Substituting the left and right ansatzes 
	\begin{eqnarray}
		W_L(z) \propto e^{-S_L(z)},\qquad W_R(z) \propto e^{-S_R(z)},
	\end{eqnarray}
	into the corresponding eigenvalue equations gives
	\begin{align}
		-4[(S_L^\prime)^2-S_L'']-i\gamma V_s\sin(z/q)&=E^*, \\
		-4[(S_R^\prime)^2-S_R'']+i\gamma V_s\sin(z/q)&=E. 
	\end{align}
	In the tail region away from the zeros of $\sin(z/q)$, the logarithmic derivatives vary slowly, such that $|S_{L,R}^{\prime\prime}|\ll|S_{L,R}^{\prime}|^2$, and the second-derivative terms can therefore be neglected.
	Choosing the branch of the square root that ensures ${\rm Re}[S_{L,R}^\prime]\ge 0$ (outward decay), we obtain
	\begin{eqnarray}
		S_{L,R}^\prime(z)\approx(1\mp i\sigma)\sqrt{\frac{\gamma V_s}{8}\left|\sin \frac{z}{q}\right|},
	\end{eqnarray}
	where $\sigma = {\rm sgn} \left[\sin (z/q)\right]$.
	Integrating outward from the corresponding Wannier centers shows that the left and right Wannier tails generally have different complex phases, $S_L\neq S_R$.
	However, their leading real decay laws are the same. 
	For $\beta=L,R$, we have
	\begin{eqnarray}
		S_\beta^{\rm (r)}(z) \approx \sqrt{\frac{\gamma V_s}{8}} \left| \int_{R_\beta}^{z} \sqrt{\left|\sin\frac{z'}{q}\right|}\,dz' \right| +o(\sqrt{\gamma}),\label{apcSr}
	\end{eqnarray}
	with $R_\beta=\pm R_c$.
	The action accumulated in the narrow matching regions near $z=n\pi q$ contributes only to subleading terms.
	The real part $S_\beta^{\rm (r)}(z)$ controls the decay of the corresponding Wannier function amplitude,
	\begin{equation}
		|W_\beta(z)|
		\approx N_\beta e^{-S_\beta^{\rm (r)}(z)},
	\end{equation}
	where $N_\beta$ is the normalization constant.
	
	Returning to the definition \eqref{hopping_eff}, the leading exponential dependence of the hopping is determined by the overlap of the two Wannier tails in the region between the neighboring centers $\pm\pi q$. Choosing $z=0$ as a convenient matching point, the overlap amplitude scales as
	\begin{eqnarray}
		\left|[W_L(-R_c)]^*W_R(R_c) \right| \sim \exp(- \kappa \sqrt{\gamma}).
	\end{eqnarray}
	Using Eq.~\eqref{apcSr}, the leading large-$\gamma$ exponent is $\kappa\sqrt{\gamma}$, where
	\begin{eqnarray}
		\kappa = q\sqrt{\frac{V_s}{8}}\int_0^{2\pi}\sqrt{|\sin u|} du, 
	\end{eqnarray}
	and the effective hopping exhibits the following behavior
	\begin{eqnarray}
		t_{\rm eff}(\gamma) \sim C({\gamma})\exp(-\kappa\sqrt{\gamma}).
	\end{eqnarray}
	The remaining operator contribution, the local matching coefficients, and the normalization of the left and right Wannier functions can be absorbed into a subexponential prefactor $C(\gamma)$, and $\ln C(\gamma) = o(\sqrt{\gamma})$.
	
	To leading order in the large-$\gamma$ regime, the logarithmic bandwidth can be approximated over a finite fitting interval by the linear form
	\begin{eqnarray}
		D_q\simeq c_1\sqrt{\gamma}+c_2,~~~~ c_1=-{\kappa}/{\ln10}<0.~~
	\end{eqnarray}
	Here, $c_2$ is an effective intercept that accounts for the subleading contribution of the prefactor $C(\gamma)$ and the local matching regions over the finite fitting interval.
	
	This $\sqrt{\gamma}$ dependence is supported by direct numerical diagonalization of the continuum model: over $2\le\gamma\le9$, the data for $D_q$ in Fig.~\ref{figAC} are well fitted by $c_1\sqrt{\gamma}+c_2$.
	The leading $\sqrt{\gamma}$ scaling of $D_q$ follows from the stretched-exponential decay of the Wannier function tails at large $\gamma$; the analytic expression for $\kappa$ qualitatively captures the asymptotic slope $c_1$.

	\bibliography{1DPTMoire.bib}

\begin{thebibliography}{79}%
\makeatletter
\providecommand \@ifxundefined [1]{%
 \@ifx{#1\undefined}
}%
\providecommand \@ifnum [1]{%
 \ifnum #1\expandafter \@firstoftwo
 \else \expandafter \@secondoftwo
 \fi
}%
\providecommand \@ifx [1]{%
 \ifx #1\expandafter \@firstoftwo
 \else \expandafter \@secondoftwo
 \fi
}%
\providecommand \natexlab [1]{#1}%
\providecommand \enquote  [1]{``#1''}%
\providecommand \bibnamefont  [1]{#1}%
\providecommand \bibfnamefont [1]{#1}%
\providecommand \citenamefont [1]{#1}%
\providecommand \href@noop [0]{\@secondoftwo}%
\providecommand \href [0]{\begingroup \@sanitize@url \@href}%
\providecommand \@href[1]{\@@startlink{#1}\@@href}%
\providecommand \@@href[1]{\endgroup#1\@@endlink}%
\providecommand \@sanitize@url [0]{\catcode `\\12\catcode `\$12\catcode
  `\&12\catcode `\#12\catcode `\^12\catcode `\_12\catcode `\%12\relax}%
\providecommand \@@startlink[1]{}%
\providecommand \@@endlink[0]{}%
\providecommand \url  [0]{\begingroup\@sanitize@url \@url }%
\providecommand \@url [1]{\endgroup\@href {#1}{\urlprefix }}%
\providecommand \urlprefix  [0]{URL }%
\providecommand \Eprint [0]{\href }%
\providecommand \doibase [0]{https://doi.org/}%
\providecommand \selectlanguage [0]{\@gobble}%
\providecommand \bibinfo  [0]{\@secondoftwo}%
\providecommand \bibfield  [0]{\@secondoftwo}%
\providecommand \translation [1]{[#1]}%
\providecommand \BibitemOpen [0]{}%
\providecommand \bibitemStop [0]{}%
\providecommand \bibitemNoStop [0]{.\EOS\space}%
\providecommand \EOS [0]{\spacefactor3000\relax}%
\providecommand \BibitemShut  [1]{\csname bibitem#1\endcsname}%
\let\auto@bib@innerbib\@empty
\bibitem [{\citenamefont {Lewenstein}\ \emph {et~al.}(2007)\citenamefont
  {Lewenstein}, \citenamefont {Sanpera}, \citenamefont {Ahufinger},
  \citenamefont {Damski}, \citenamefont {Sen(De)},\ and\ \citenamefont
  {Sen}}]{LewensteinSen2007AiP}%
  \BibitemOpen
  \bibfield  {author} {\bibinfo {author} {\bibfnamefont {M.}~\bibnamefont
  {Lewenstein}}, \bibinfo {author} {\bibfnamefont {A.}~\bibnamefont {Sanpera}},
  \bibinfo {author} {\bibfnamefont {V.}~\bibnamefont {Ahufinger}}, \bibinfo
  {author} {\bibfnamefont {B.}~\bibnamefont {Damski}}, \bibinfo {author}
  {\bibfnamefont {A.}~\bibnamefont {Sen(De)}},\ and\ \bibinfo {author}
  {\bibfnamefont {U.}~\bibnamefont {Sen}},\ }\href
  {https://doi.org/10.1080/00018730701223200} {\bibfield  {journal} {\bibinfo
  {journal} {Adv. Phys.}\ }\textbf {\bibinfo {volume} {56}},\ \bibinfo {pages}
  {243} (\bibinfo {year} {2007})}\BibitemShut {NoStop}%
\bibitem [{\citenamefont {Bloch}\ \emph {et~al.}(2008)\citenamefont {Bloch},
  \citenamefont {Dalibard},\ and\ \citenamefont
  {Zwerger}}]{BlochZwerger2008RMP}%
  \BibitemOpen
  \bibfield  {author} {\bibinfo {author} {\bibfnamefont {I.}~\bibnamefont
  {Bloch}}, \bibinfo {author} {\bibfnamefont {J.}~\bibnamefont {Dalibard}},\
  and\ \bibinfo {author} {\bibfnamefont {W.}~\bibnamefont {Zwerger}},\ }\href
  {https://doi.org/10.1103/RevModPhys.80.885} {\bibfield  {journal} {\bibinfo
  {journal} {Rev. Mod. Phys.}\ }\textbf {\bibinfo {volume} {80}},\ \bibinfo
  {pages} {885} (\bibinfo {year} {2008})}\BibitemShut {NoStop}%
\bibitem [{\citenamefont {Zhang}\ \emph {et~al.}(2018)\citenamefont {Zhang},
  \citenamefont {Zhu}, \citenamefont {Zhao}, \citenamefont {Yan},\ and\
  \citenamefont {Zhu}}]{ZhangZhu2018AiP}%
  \BibitemOpen
  \bibfield  {author} {\bibinfo {author} {\bibfnamefont {D.-W.}\ \bibnamefont
  {Zhang}}, \bibinfo {author} {\bibfnamefont {Y.-Q.}\ \bibnamefont {Zhu}},
  \bibinfo {author} {\bibfnamefont {Y.~X.}\ \bibnamefont {Zhao}}, \bibinfo
  {author} {\bibfnamefont {H.}~\bibnamefont {Yan}},\ and\ \bibinfo {author}
  {\bibfnamefont {S.-L.}\ \bibnamefont {Zhu}},\ }\href
  {https://doi.org/10.1080/00018732.2019.1594094} {\bibfield  {journal}
  {\bibinfo  {journal} {Adv. Phys.}\ }\textbf {\bibinfo {volume} {67}},\
  \bibinfo {pages} {253} (\bibinfo {year} {2018})}\BibitemShut {NoStop}%
\bibitem [{\citenamefont {Meng}\ \emph {et~al.}(2023)\citenamefont {Meng},
  \citenamefont {Wang}, \citenamefont {Han}, \citenamefont {Liu}, \citenamefont
  {Wen}, \citenamefont {Gao}, \citenamefont {Wang}, \citenamefont {Chin},\ and\
  \citenamefont {Zhang}}]{MengZhang2023Nature}%
  \BibitemOpen
  \bibfield  {author} {\bibinfo {author} {\bibfnamefont {Z.}~\bibnamefont
  {Meng}}, \bibinfo {author} {\bibfnamefont {L.}~\bibnamefont {Wang}}, \bibinfo
  {author} {\bibfnamefont {W.}~\bibnamefont {Han}}, \bibinfo {author}
  {\bibfnamefont {F.}~\bibnamefont {Liu}}, \bibinfo {author} {\bibfnamefont
  {K.}~\bibnamefont {Wen}}, \bibinfo {author} {\bibfnamefont {C.}~\bibnamefont
  {Gao}}, \bibinfo {author} {\bibfnamefont {P.}~\bibnamefont {Wang}}, \bibinfo
  {author} {\bibfnamefont {C.}~\bibnamefont {Chin}},\ and\ \bibinfo {author}
  {\bibfnamefont {J.}~\bibnamefont {Zhang}},\ }\href
  {https://doi.org/10.1038/s41586-023-05695-4} {\bibfield  {journal} {\bibinfo
  {journal} {Nature}\ }\textbf {\bibinfo {volume} {615}},\ \bibinfo {pages}
  {231} (\bibinfo {year} {2023})}\BibitemShut {NoStop}%
\bibitem [{\citenamefont {Cao}\ \emph {et~al.}(2018{\natexlab{a}})\citenamefont
  {Cao}, \citenamefont {Fatemi}, \citenamefont {Fang}, \citenamefont
  {Watanabe}, \citenamefont {Taniguchi}, \citenamefont {Kaxiras},\ and\
  \citenamefont {Jarillo-Herrero}}]{CaoJarillo-Herrero2018Nature}%
  \BibitemOpen
  \bibfield  {author} {\bibinfo {author} {\bibfnamefont {Y.}~\bibnamefont
  {Cao}}, \bibinfo {author} {\bibfnamefont {V.}~\bibnamefont {Fatemi}},
  \bibinfo {author} {\bibfnamefont {S.}~\bibnamefont {Fang}}, \bibinfo {author}
  {\bibfnamefont {K.}~\bibnamefont {Watanabe}}, \bibinfo {author}
  {\bibfnamefont {T.}~\bibnamefont {Taniguchi}}, \bibinfo {author}
  {\bibfnamefont {E.}~\bibnamefont {Kaxiras}},\ and\ \bibinfo {author}
  {\bibfnamefont {P.}~\bibnamefont {Jarillo-Herrero}},\ }\href
  {https://doi.org/10.1038/nature26160} {\bibfield  {journal} {\bibinfo
  {journal} {Nature}\ }\textbf {\bibinfo {volume} {556}},\ \bibinfo {pages}
  {43} (\bibinfo {year} {2018}{\natexlab{a}})}\BibitemShut {NoStop}%
\bibitem [{\citenamefont {Cao}\ \emph {et~al.}(2018{\natexlab{b}})\citenamefont
  {Cao}, \citenamefont {Fatemi}, \citenamefont {Demir}, \citenamefont {Fang},
  \citenamefont {Tomarken}, \citenamefont {Luo}, \citenamefont
  {Sanchez-Yamagishi}, \citenamefont {Watanabe}, \citenamefont {Taniguchi},\
  and\ \citenamefont {Kaxiras}}]{CaoKaxiras2018Nature}%
  \BibitemOpen
  \bibfield  {author} {\bibinfo {author} {\bibfnamefont {Y.}~\bibnamefont
  {Cao}}, \bibinfo {author} {\bibfnamefont {V.}~\bibnamefont {Fatemi}},
  \bibinfo {author} {\bibfnamefont {A.}~\bibnamefont {Demir}}, \bibinfo
  {author} {\bibfnamefont {S.}~\bibnamefont {Fang}}, \bibinfo {author}
  {\bibfnamefont {S.~L.}\ \bibnamefont {Tomarken}}, \bibinfo {author}
  {\bibfnamefont {J.~Y.}\ \bibnamefont {Luo}}, \bibinfo {author} {\bibfnamefont
  {J.~D.}\ \bibnamefont {Sanchez-Yamagishi}}, \bibinfo {author} {\bibfnamefont
  {K.}~\bibnamefont {Watanabe}}, \bibinfo {author} {\bibfnamefont
  {T.}~\bibnamefont {Taniguchi}},\ and\ \bibinfo {author} {\bibfnamefont
  {E.}~\bibnamefont {Kaxiras}},\ }\href {https://doi.org/10.1038/nature26154}
  {\bibfield  {journal} {\bibinfo  {journal} {Nature}\ }\textbf {\bibinfo
  {volume} {556}},\ \bibinfo {pages} {80} (\bibinfo {year}
  {2018}{\natexlab{b}})}\BibitemShut {NoStop}%
\bibitem [{\citenamefont {Lu}\ \emph {et~al.}(2019)\citenamefont {Lu},
  \citenamefont {Stepanov}, \citenamefont {Yang}, \citenamefont {Xie},
  \citenamefont {Aamir}, \citenamefont {Das}, \citenamefont {Urgell},
  \citenamefont {Watanabe}, \citenamefont {Taniguchi}, \citenamefont {Zhang},
  \citenamefont {Bachtold}, \citenamefont {MacDonald},\ and\ \citenamefont
  {Efetov}}]{LuDmitri2019nature}%
  \BibitemOpen
  \bibfield  {author} {\bibinfo {author} {\bibfnamefont {X.}~\bibnamefont
  {Lu}}, \bibinfo {author} {\bibfnamefont {P.}~\bibnamefont {Stepanov}},
  \bibinfo {author} {\bibfnamefont {W.}~\bibnamefont {Yang}}, \bibinfo {author}
  {\bibfnamefont {M.}~\bibnamefont {Xie}}, \bibinfo {author} {\bibfnamefont
  {M.~A.}\ \bibnamefont {Aamir}}, \bibinfo {author} {\bibfnamefont
  {I.}~\bibnamefont {Das}}, \bibinfo {author} {\bibfnamefont {C.}~\bibnamefont
  {Urgell}}, \bibinfo {author} {\bibfnamefont {K.}~\bibnamefont {Watanabe}},
  \bibinfo {author} {\bibfnamefont {T.}~\bibnamefont {Taniguchi}}, \bibinfo
  {author} {\bibfnamefont {G.}~\bibnamefont {Zhang}}, \bibinfo {author}
  {\bibfnamefont {A.}~\bibnamefont {Bachtold}}, \bibinfo {author}
  {\bibfnamefont {A.~H.}\ \bibnamefont {MacDonald}},\ and\ \bibinfo {author}
  {\bibfnamefont {D.~K.}\ \bibnamefont {Efetov}},\ }\href
  {https://doi.org/10.1038/s41586-019-1695-0} {\bibfield  {journal} {\bibinfo
  {journal} {Nature}\ }\textbf {\bibinfo {volume} {574}},\ \bibinfo {pages}
  {653} (\bibinfo {year} {2019})}\BibitemShut {NoStop}%
\bibitem [{\citenamefont {Yankowitz}\ \emph {et~al.}(2019)\citenamefont
  {Yankowitz}, \citenamefont {Chen}, \citenamefont {Polshyn}, \citenamefont
  {Zhang}, \citenamefont {Watanabe}, \citenamefont {Taniguchi}, \citenamefont
  {Graf}, \citenamefont {Young},\ and\ \citenamefont
  {Dean}}]{Yankowitz2019Science}%
  \BibitemOpen
  \bibfield  {author} {\bibinfo {author} {\bibfnamefont {M.}~\bibnamefont
  {Yankowitz}}, \bibinfo {author} {\bibfnamefont {S.}~\bibnamefont {Chen}},
  \bibinfo {author} {\bibfnamefont {H.}~\bibnamefont {Polshyn}}, \bibinfo
  {author} {\bibfnamefont {Y.}~\bibnamefont {Zhang}}, \bibinfo {author}
  {\bibfnamefont {K.}~\bibnamefont {Watanabe}}, \bibinfo {author}
  {\bibfnamefont {T.}~\bibnamefont {Taniguchi}}, \bibinfo {author}
  {\bibfnamefont {D.}~\bibnamefont {Graf}}, \bibinfo {author} {\bibfnamefont
  {A.~F.}\ \bibnamefont {Young}},\ and\ \bibinfo {author} {\bibfnamefont
  {C.~R.}\ \bibnamefont {Dean}},\ }\href
  {https://doi.org/10.1126/science.aav1910} {\bibfield  {journal} {\bibinfo
  {journal} {Science}\ }\textbf {\bibinfo {volume} {363}},\ \bibinfo {pages}
  {1059} (\bibinfo {year} {2019})}\BibitemShut {NoStop}%
\bibitem [{\citenamefont {Jin}\ \emph {et~al.}(2021)\citenamefont {Jin},
  \citenamefont {Tao}, \citenamefont {Li}, \citenamefont {Xu}, \citenamefont
  {Tang}, \citenamefont {Zhu}, \citenamefont {Liu}, \citenamefont {Watanabe},
  \citenamefont {Taniguchi}, \citenamefont {Hone}, \citenamefont {Fu},
  \citenamefont {Shan},\ and\ \citenamefont {Mak}}]{JinKin2021nature}%
  \BibitemOpen
  \bibfield  {author} {\bibinfo {author} {\bibfnamefont {C.}~\bibnamefont
  {Jin}}, \bibinfo {author} {\bibfnamefont {Z.}~\bibnamefont {Tao}}, \bibinfo
  {author} {\bibfnamefont {T.}~\bibnamefont {Li}}, \bibinfo {author}
  {\bibfnamefont {Y.}~\bibnamefont {Xu}}, \bibinfo {author} {\bibfnamefont
  {Y.}~\bibnamefont {Tang}}, \bibinfo {author} {\bibfnamefont {J.}~\bibnamefont
  {Zhu}}, \bibinfo {author} {\bibfnamefont {S.}~\bibnamefont {Liu}}, \bibinfo
  {author} {\bibfnamefont {K.}~\bibnamefont {Watanabe}}, \bibinfo {author}
  {\bibfnamefont {T.}~\bibnamefont {Taniguchi}}, \bibinfo {author}
  {\bibfnamefont {J.~C.}\ \bibnamefont {Hone}}, \bibinfo {author}
  {\bibfnamefont {L.}~\bibnamefont {Fu}}, \bibinfo {author} {\bibfnamefont
  {J.}~\bibnamefont {Shan}},\ and\ \bibinfo {author} {\bibfnamefont {K.~F.}\
  \bibnamefont {Mak}},\ }\href {https://doi.org/10.1038/s41563-021-00959-8}
  {\bibfield  {journal} {\bibinfo  {journal} {Nat. Mater.}\ }\textbf {\bibinfo
  {volume} {20}},\ \bibinfo {pages} {940} (\bibinfo {year} {2021})}\BibitemShut
  {NoStop}%
\bibitem [{\citenamefont {Vu}\ and\ \citenamefont {Das~Sarma}(2021)}]{Vu2021}%
  \BibitemOpen
  \bibfield  {author} {\bibinfo {author} {\bibfnamefont {D.}~\bibnamefont
  {Vu}}\ and\ \bibinfo {author} {\bibfnamefont {S.}~\bibnamefont {Das~Sarma}},\
  }\href {https://link.aps.org/doi/10.1103/PhysRevLett.126.036803} {\bibfield
  {journal} {\bibinfo  {journal} {Phys. Rev. Lett.}\ }\textbf {\bibinfo
  {volume} {126}},\ \bibinfo {pages} {036803} (\bibinfo {year}
  {2021})}\BibitemShut {NoStop}%
\bibitem [{\citenamefont {Gonçalves}\ \emph {et~al.}(2024)\citenamefont
  {Gonçalves}, \citenamefont {Amorim}, \citenamefont {Riche}, \citenamefont
  {Castro},\ and\ \citenamefont {Ribeiro}}]{Goncalves2024NP}%
  \BibitemOpen
  \bibfield  {author} {\bibinfo {author} {\bibfnamefont {M.}~\bibnamefont
  {Gonçalves}}, \bibinfo {author} {\bibfnamefont {B.}~\bibnamefont {Amorim}},
  \bibinfo {author} {\bibfnamefont {F.}~\bibnamefont {Riche}}, \bibinfo
  {author} {\bibfnamefont {E.~V.}\ \bibnamefont {Castro}},\ and\ \bibinfo
  {author} {\bibfnamefont {P.}~\bibnamefont {Ribeiro}},\ }\href
  {https://doi.org/10.1038/s41567-024-02662-2} {\bibfield  {journal} {\bibinfo
  {journal} {Nat. Phys.}\ }\textbf {\bibinfo {volume} {20}},\ \bibinfo {pages}
  {1933} (\bibinfo {year} {2024})}\BibitemShut {NoStop}%
\bibitem [{\citenamefont {Zhang}\ \emph {et~al.}(2025)\citenamefont {Zhang},
  \citenamefont {Tang}, \citenamefont {Quezada}, \citenamefont {Dong},\ and\
  \citenamefont {Zhang}}]{ZhangZhang2025CommPhy}%
  \BibitemOpen
  \bibfield  {author} {\bibinfo {author} {\bibfnamefont {G.-Q.}\ \bibnamefont
  {Zhang}}, \bibinfo {author} {\bibfnamefont {L.-Z.}\ \bibnamefont {Tang}},
  \bibinfo {author} {\bibfnamefont {L.~F.}\ \bibnamefont {Quezada}}, \bibinfo
  {author} {\bibfnamefont {S.-H.}\ \bibnamefont {Dong}},\ and\ \bibinfo
  {author} {\bibfnamefont {D.-W.}\ \bibnamefont {Zhang}},\ }\href
  {https://doi.org/10.1038/s42005-025-02197-9} {\bibfield  {journal} {\bibinfo
  {journal} {Commun. Phys.}\ }\textbf {\bibinfo {volume} {8}} (\bibinfo {year}
  {2025})}\BibitemShut {NoStop}%
\bibitem [{\citenamefont {Nguyen}\ \emph {et~al.}(2022)\citenamefont {Nguyen},
  \citenamefont {Letartre}, \citenamefont {Drouard}, \citenamefont
  {Viktorovitch}, \citenamefont {Nguyen},\ and\ \citenamefont
  {Nguyen}}]{NguyenNguyen2022PRR}%
  \BibitemOpen
  \bibfield  {author} {\bibinfo {author} {\bibfnamefont {D.~X.}\ \bibnamefont
  {Nguyen}}, \bibinfo {author} {\bibfnamefont {X.}~\bibnamefont {Letartre}},
  \bibinfo {author} {\bibfnamefont {E.}~\bibnamefont {Drouard}}, \bibinfo
  {author} {\bibfnamefont {P.}~\bibnamefont {Viktorovitch}}, \bibinfo {author}
  {\bibfnamefont {H.~C.}\ \bibnamefont {Nguyen}},\ and\ \bibinfo {author}
  {\bibfnamefont {H.~S.}\ \bibnamefont {Nguyen}},\ }\href
  {https://doi.org/10.1103/PhysRevResearch.4.L032031} {\bibfield  {journal}
  {\bibinfo  {journal} {Phys. Rev. Res.}\ }\textbf {\bibinfo {volume} {4}},\
  \bibinfo {pages} {L032031} (\bibinfo {year} {2022})}\BibitemShut {NoStop}%
\bibitem [{\citenamefont {Talukdar}\ \emph {et~al.}(2022)\citenamefont
  {Talukdar}, \citenamefont {Hardison},\ and\ \citenamefont
  {Ryckman}}]{TalukdarRyckman2022ACSPho}%
  \BibitemOpen
  \bibfield  {author} {\bibinfo {author} {\bibfnamefont {T.~H.}\ \bibnamefont
  {Talukdar}}, \bibinfo {author} {\bibfnamefont {A.~L.}\ \bibnamefont
  {Hardison}},\ and\ \bibinfo {author} {\bibfnamefont {J.~D.}\ \bibnamefont
  {Ryckman}},\ }\href {https://doi.org/10.1021/acsphotonics.1c01800} {\bibfield
   {journal} {\bibinfo  {journal} {ACS Photonics}\ }\textbf {\bibinfo {volume}
  {9}},\ \bibinfo {pages} {1286} (\bibinfo {year} {2022})}\BibitemShut
  {NoStop}%
\bibitem [{\citenamefont {Yu}\ \emph {et~al.}(2023)\citenamefont {Yu},
  \citenamefont {Li}, \citenamefont {Wang}, \citenamefont {Leykam},
  \citenamefont {Yuan},\ and\ \citenamefont {Chen}}]{YuChen2023PRL}%
  \BibitemOpen
  \bibfield  {author} {\bibinfo {author} {\bibfnamefont {D.}~\bibnamefont
  {Yu}}, \bibinfo {author} {\bibfnamefont {G.}~\bibnamefont {Li}}, \bibinfo
  {author} {\bibfnamefont {L.}~\bibnamefont {Wang}}, \bibinfo {author}
  {\bibfnamefont {D.}~\bibnamefont {Leykam}}, \bibinfo {author} {\bibfnamefont
  {L.}~\bibnamefont {Yuan}},\ and\ \bibinfo {author} {\bibfnamefont
  {X.}~\bibnamefont {Chen}},\ }\href
  {https://doi.org/10.1103/physrevlett.130.143801} {\bibfield  {journal}
  {\bibinfo  {journal} {Phys. Rev. Lett.}\ }\textbf {\bibinfo {volume} {130}},\
  \bibinfo {pages} {143801} (\bibinfo {year} {2023})}\BibitemShut {NoStop}%
\bibitem [{\citenamefont {Xia}\ \emph {et~al.}(2024)\citenamefont {Xia},
  \citenamefont {Liu}, \citenamefont {Zou}, \citenamefont {Hong},\ and\
  \citenamefont {Liang}}]{XiaLiang2024OL}%
  \BibitemOpen
  \bibfield  {author} {\bibinfo {author} {\bibfnamefont {X.}~\bibnamefont
  {Xia}}, \bibinfo {author} {\bibfnamefont {Q.}~\bibnamefont {Liu}}, \bibinfo
  {author} {\bibfnamefont {B.}~\bibnamefont {Zou}}, \bibinfo {author}
  {\bibfnamefont {P.}~\bibnamefont {Hong}},\ and\ \bibinfo {author}
  {\bibfnamefont {Y.}~\bibnamefont {Liang}},\ }\href
  {https://doi.org/10.1364/OL.522215} {\bibfield  {journal} {\bibinfo
  {journal} {Opt. Lett.}\ }\textbf {\bibinfo {volume} {49}},\ \bibinfo {pages}
  {2553} (\bibinfo {year} {2024})}\BibitemShut {NoStop}%
\bibitem [{\citenamefont {Trushin}\ \emph {et~al.}(2025)\citenamefont
  {Trushin}, \citenamefont {Ito}, \citenamefont {Ishii}, \citenamefont
  {Iwamoto},\ and\ \citenamefont {Ota}}]{TrushinOta2025OL}%
  \BibitemOpen
  \bibfield  {author} {\bibinfo {author} {\bibfnamefont {S.~M.}\ \bibnamefont
  {Trushin}}, \bibinfo {author} {\bibfnamefont {T.}~\bibnamefont {Ito}},
  \bibinfo {author} {\bibfnamefont {Y.}~\bibnamefont {Ishii}}, \bibinfo
  {author} {\bibfnamefont {S.}~\bibnamefont {Iwamoto}},\ and\ \bibinfo {author}
  {\bibfnamefont {Y.}~\bibnamefont {Ota}},\ }\href
  {https://doi.org/10.1364/OL.558564} {\bibfield  {journal} {\bibinfo
  {journal} {Opt. Lett.}\ }\textbf {\bibinfo {volume} {50}},\ \bibinfo {pages}
  {2405} (\bibinfo {year} {2025})}\BibitemShut {NoStop}%
\bibitem [{\citenamefont {Li}\ \emph {et~al.}(2025)\citenamefont {Li},
  \citenamefont {He}, \citenamefont {Wang}, \citenamefont {Yang}, \citenamefont
  {Yu}, \citenamefont {Zheng}, \citenamefont {Yuan},\ and\ \citenamefont
  {Chen}}]{LiChen2025PRL}%
  \BibitemOpen
  \bibfield  {author} {\bibinfo {author} {\bibfnamefont {G.}~\bibnamefont
  {Li}}, \bibinfo {author} {\bibfnamefont {Y.}~\bibnamefont {He}}, \bibinfo
  {author} {\bibfnamefont {L.}~\bibnamefont {Wang}}, \bibinfo {author}
  {\bibfnamefont {Y.}~\bibnamefont {Yang}}, \bibinfo {author} {\bibfnamefont
  {D.}~\bibnamefont {Yu}}, \bibinfo {author} {\bibfnamefont {Y.}~\bibnamefont
  {Zheng}}, \bibinfo {author} {\bibfnamefont {L.}~\bibnamefont {Yuan}},\ and\
  \bibinfo {author} {\bibfnamefont {X.}~\bibnamefont {Chen}},\ }\href
  {https://doi.org/10.1103/PhysRevLett.134.083803} {\bibfield  {journal}
  {\bibinfo  {journal} {Phys. Rev. Lett.}\ }\textbf {\bibinfo {volume} {134}},\
  \bibinfo {pages} {083803} (\bibinfo {year} {2025})}\BibitemShut {NoStop}%
\bibitem [{\citenamefont {Nath}\ and\ \citenamefont
  {Roy}(2014)}]{NathRoy2014LPL}%
  \BibitemOpen
  \bibfield  {author} {\bibinfo {author} {\bibfnamefont {A.}~\bibnamefont
  {Nath}}\ and\ \bibinfo {author} {\bibfnamefont {U.}~\bibnamefont {Roy}},\
  }\href {https://doi.org/10.1088/1612-2011/11/11/115501} {\bibfield  {journal}
  {\bibinfo  {journal} {Laser Phys. Lett.}\ }\textbf {\bibinfo {volume} {11}},\
  \bibinfo {pages} {115501} (\bibinfo {year} {2014})}\BibitemShut {NoStop}%
\bibitem [{\citenamefont {Nath}\ \emph {et~al.}(2022)\citenamefont {Nath},
  \citenamefont {Bera}, \citenamefont {Pathak},\ and\ \citenamefont
  {Roy}}]{NathRoy2022TEPJD}%
  \BibitemOpen
  \bibfield  {author} {\bibinfo {author} {\bibfnamefont {A.}~\bibnamefont
  {Nath}}, \bibinfo {author} {\bibfnamefont {J.}~\bibnamefont {Bera}}, \bibinfo
  {author} {\bibfnamefont {M.~R.}\ \bibnamefont {Pathak}},\ and\ \bibinfo
  {author} {\bibfnamefont {U.}~\bibnamefont {Roy}},\ }\href
  {https://doi.org/10.1140/epjd/s10053-022-00571-8} {\bibfield  {journal}
  {\bibinfo  {journal} {EUR PHYS J D}\ }\textbf {\bibinfo {volume} {76}}
  (\bibinfo {year} {2022})}\BibitemShut {NoStop}%
\bibitem [{\citenamefont {Raghav}\ \emph {et~al.}(2022)\citenamefont {Raghav},
  \citenamefont {Halder}, \citenamefont {Basu},\ and\ \citenamefont
  {Roy}}]{RaghavRoy2022PRA}%
  \BibitemOpen
  \bibfield  {author} {\bibinfo {author} {\bibfnamefont {S.}~\bibnamefont
  {Raghav}}, \bibinfo {author} {\bibfnamefont {B.}~\bibnamefont {Halder}},
  \bibinfo {author} {\bibfnamefont {P.}~\bibnamefont {Basu}},\ and\ \bibinfo
  {author} {\bibfnamefont {U.}~\bibnamefont {Roy}},\ }\href
  {https://doi.org/10.1103/PhysRevA.106.063304} {\bibfield  {journal} {\bibinfo
   {journal} {Phys. Rev. A}\ }\textbf {\bibinfo {volume} {106}},\ \bibinfo
  {pages} {063304} (\bibinfo {year} {2022})}\BibitemShut {NoStop}%
\bibitem [{\citenamefont {Zhou}\ \emph {et~al.}(2025)\citenamefont {Zhou},
  \citenamefont {Li}, \citenamefont {Zhang}, \citenamefont {Lan}, \citenamefont
  {Celi},\ and\ \citenamefont {Zhang}}]{ZhouZhang2025PRA}%
  \BibitemOpen
  \bibfield  {author} {\bibinfo {author} {\bibfnamefont {L.}~\bibnamefont
  {Zhou}}, \bibinfo {author} {\bibfnamefont {Z.-C.}\ \bibnamefont {Li}},
  \bibinfo {author} {\bibfnamefont {K.}~\bibnamefont {Zhang}}, \bibinfo
  {author} {\bibfnamefont {Z.}~\bibnamefont {Lan}}, \bibinfo {author}
  {\bibfnamefont {A.}~\bibnamefont {Celi}},\ and\ \bibinfo {author}
  {\bibfnamefont {W.}~\bibnamefont {Zhang}},\ }\href
  {https://doi.org/10.1103/6751-zclb} {\bibfield  {journal} {\bibinfo
  {journal} {Phys. Rev. A}\ }\textbf {\bibinfo {volume} {112}},\ \bibinfo
  {pages} {043718} (\bibinfo {year} {2025})}\BibitemShut {NoStop}%
\bibitem [{\citenamefont {Roati}\ \emph {et~al.}(2008)\citenamefont {Roati},
  \citenamefont {D'Errico}, \citenamefont {Fallani}, \citenamefont {Fattori},
  \citenamefont {Fort}, \citenamefont {Zaccanti}, \citenamefont {Modugno},
  \citenamefont {Modugno},\ and\ \citenamefont {Inguscio}}]{Roati2008Roati}%
  \BibitemOpen
  \bibfield  {author} {\bibinfo {author} {\bibfnamefont {G.}~\bibnamefont
  {Roati}}, \bibinfo {author} {\bibfnamefont {C.}~\bibnamefont {D'Errico}},
  \bibinfo {author} {\bibfnamefont {L.}~\bibnamefont {Fallani}}, \bibinfo
  {author} {\bibfnamefont {M.}~\bibnamefont {Fattori}}, \bibinfo {author}
  {\bibfnamefont {C.}~\bibnamefont {Fort}}, \bibinfo {author} {\bibfnamefont
  {M.}~\bibnamefont {Zaccanti}}, \bibinfo {author} {\bibfnamefont
  {G.}~\bibnamefont {Modugno}}, \bibinfo {author} {\bibfnamefont
  {M.}~\bibnamefont {Modugno}},\ and\ \bibinfo {author} {\bibfnamefont
  {M.}~\bibnamefont {Inguscio}},\ }\href {https://doi.org/10.1038/nature07071}
  {\bibfield  {journal} {\bibinfo  {journal} {Nature}\ }\textbf {\bibinfo
  {volume} {453}},\ \bibinfo {pages} {895} (\bibinfo {year}
  {2008})}\BibitemShut {NoStop}%
\bibitem [{\citenamefont {Schreiber}\ \emph {et~al.}(2015)\citenamefont
  {Schreiber}, \citenamefont {Hodgman}, \citenamefont {Bordia}, \citenamefont
  {Lüschen}, \citenamefont {Fischer}, \citenamefont {Vosk}, \citenamefont
  {Altman}, \citenamefont {Schneider},\ and\ \citenamefont
  {Bloch}}]{Schreiber2015Science}%
  \BibitemOpen
  \bibfield  {author} {\bibinfo {author} {\bibfnamefont {M.}~\bibnamefont
  {Schreiber}}, \bibinfo {author} {\bibfnamefont {S.~S.}\ \bibnamefont
  {Hodgman}}, \bibinfo {author} {\bibfnamefont {P.}~\bibnamefont {Bordia}},
  \bibinfo {author} {\bibfnamefont {H.~P.}\ \bibnamefont {Lüschen}}, \bibinfo
  {author} {\bibfnamefont {M.~H.}\ \bibnamefont {Fischer}}, \bibinfo {author}
  {\bibfnamefont {R.}~\bibnamefont {Vosk}}, \bibinfo {author} {\bibfnamefont
  {E.}~\bibnamefont {Altman}}, \bibinfo {author} {\bibfnamefont
  {U.}~\bibnamefont {Schneider}},\ and\ \bibinfo {author} {\bibfnamefont
  {I.}~\bibnamefont {Bloch}},\ }\href {https://doi.org/10.1126/science.aaa7432}
  {\bibfield  {journal} {\bibinfo  {journal} {Science}\ }\textbf {\bibinfo
  {volume} {349}},\ \bibinfo {pages} {842} (\bibinfo {year}
  {2015})}\BibitemShut {NoStop}%
\bibitem [{\citenamefont {Kohlert}\ \emph {et~al.}(2019)\citenamefont
  {Kohlert}, \citenamefont {Scherg}, \citenamefont {Li}, \citenamefont
  {L\"uschen}, \citenamefont {Das~Sarma}, \citenamefont {Bloch},\ and\
  \citenamefont {Aidelsburger}}]{KohertAidelsburger2019PRL}%
  \BibitemOpen
  \bibfield  {author} {\bibinfo {author} {\bibfnamefont {T.}~\bibnamefont
  {Kohlert}}, \bibinfo {author} {\bibfnamefont {S.}~\bibnamefont {Scherg}},
  \bibinfo {author} {\bibfnamefont {X.}~\bibnamefont {Li}}, \bibinfo {author}
  {\bibfnamefont {H.~P.}\ \bibnamefont {L\"uschen}}, \bibinfo {author}
  {\bibfnamefont {S.}~\bibnamefont {Das~Sarma}}, \bibinfo {author}
  {\bibfnamefont {I.}~\bibnamefont {Bloch}},\ and\ \bibinfo {author}
  {\bibfnamefont {M.}~\bibnamefont {Aidelsburger}},\ }\href
  {https://doi.org/10.1103/PhysRevLett.122.170403} {\bibfield  {journal}
  {\bibinfo  {journal} {Phys. Rev. Lett.}\ }\textbf {\bibinfo {volume} {122}},\
  \bibinfo {pages} {170403} (\bibinfo {year} {2019})}\BibitemShut {NoStop}%
\bibitem [{\citenamefont {Ashida}\ \emph {et~al.}(2020)\citenamefont {Ashida},
  \citenamefont {Gong},\ and\ \citenamefont {Ueda}}]{AshidaUeda2020}%
  \BibitemOpen
  \bibfield  {author} {\bibinfo {author} {\bibfnamefont {Y.}~\bibnamefont
  {Ashida}}, \bibinfo {author} {\bibfnamefont {Z.}~\bibnamefont {Gong}},\ and\
  \bibinfo {author} {\bibfnamefont {M.}~\bibnamefont {Ueda}},\ }\href
  {https://doi.org/10.1080/00018732.2021.1876991} {\bibfield  {journal}
  {\bibinfo  {journal} {Adv. Phys.}\ }\textbf {\bibinfo {volume} {69}},\
  \bibinfo {pages} {249} (\bibinfo {year} {2020})}\BibitemShut {NoStop}%
\bibitem [{\citenamefont {Bergholtz}\ \emph {et~al.}(2021)\citenamefont
  {Bergholtz}, \citenamefont {Budich},\ and\ \citenamefont
  {Kunst}}]{BergholtzKunst2021}%
  \BibitemOpen
  \bibfield  {author} {\bibinfo {author} {\bibfnamefont {E.~J.}\ \bibnamefont
  {Bergholtz}}, \bibinfo {author} {\bibfnamefont {J.~C.}\ \bibnamefont
  {Budich}},\ and\ \bibinfo {author} {\bibfnamefont {F.~K.}\ \bibnamefont
  {Kunst}},\ }\href {https://doi.org/10.1103/RevModPhys.93.015005} {\bibfield
  {journal} {\bibinfo  {journal} {Rev. Mod. Phys.}\ }\textbf {\bibinfo {volume}
  {93}},\ \bibinfo {pages} {015005} (\bibinfo {year} {2021})}\BibitemShut
  {NoStop}%
\bibitem [{\citenamefont {Bender}\ and\ \citenamefont
  {Boettcher}(1998)}]{BenderBoettcher1998PRL}%
  \BibitemOpen
  \bibfield  {author} {\bibinfo {author} {\bibfnamefont {C.~M.}\ \bibnamefont
  {Bender}}\ and\ \bibinfo {author} {\bibfnamefont {S.}~\bibnamefont
  {Boettcher}},\ }\href {https://doi.org/10.1103/PhysRevLett.80.5243}
  {\bibfield  {journal} {\bibinfo  {journal} {Phys. Rev. Lett.}\ }\textbf
  {\bibinfo {volume} {80}},\ \bibinfo {pages} {5243} (\bibinfo {year}
  {1998})}\BibitemShut {NoStop}%
\bibitem [{\citenamefont {Bender}(2007)}]{Bender2007}%
  \BibitemOpen
  \bibfield  {author} {\bibinfo {author} {\bibfnamefont {C.~M.}\ \bibnamefont
  {Bender}},\ }\href {https://doi.org/10.1088/0034-4885/70/6/R03} {\bibfield
  {journal} {\bibinfo  {journal} {Rep. Prog. Phys.}\ }\textbf {\bibinfo
  {volume} {70}},\ \bibinfo {pages} {947} (\bibinfo {year} {2007})}\BibitemShut
  {NoStop}%
\bibitem [{\citenamefont {Konotop}\ \emph {et~al.}(2016)\citenamefont
  {Konotop}, \citenamefont {Yang},\ and\ \citenamefont
  {Zezyulin}}]{KonotopDmitry2016RMP}%
  \BibitemOpen
  \bibfield  {author} {\bibinfo {author} {\bibfnamefont {V.~V.}\ \bibnamefont
  {Konotop}}, \bibinfo {author} {\bibfnamefont {J.}~\bibnamefont {Yang}},\ and\
  \bibinfo {author} {\bibfnamefont {D.~A.}\ \bibnamefont {Zezyulin}},\ }\href
  {https://doi.org/10.1103/RevModPhys.88.035002} {\bibfield  {journal}
  {\bibinfo  {journal} {Rev. Mod. Phys.}\ }\textbf {\bibinfo {volume} {88}},\
  \bibinfo {pages} {035002} (\bibinfo {year} {2016})}\BibitemShut {NoStop}%
\bibitem [{\citenamefont {Musslimani}\ \emph {et~al.}(2008)\citenamefont
  {Musslimani}, \citenamefont {Makris}, \citenamefont {El-Ganainy},\ and\
  \citenamefont {Christodoulides}}]{MusslimaniChristodoulides2008PRL}%
  \BibitemOpen
  \bibfield  {author} {\bibinfo {author} {\bibfnamefont {Z.~H.}\ \bibnamefont
  {Musslimani}}, \bibinfo {author} {\bibfnamefont {K.~G.}\ \bibnamefont
  {Makris}}, \bibinfo {author} {\bibfnamefont {R.}~\bibnamefont {El-Ganainy}},\
  and\ \bibinfo {author} {\bibfnamefont {D.~N.}\ \bibnamefont
  {Christodoulides}},\ }\href {https://doi.org/10.1103/PhysRevLett.100.030402}
  {\bibfield  {journal} {\bibinfo  {journal} {Phys. Rev. Lett.}\ }\textbf
  {\bibinfo {volume} {100}},\ \bibinfo {pages} {030402} (\bibinfo {year}
  {2008})}\BibitemShut {NoStop}%
\bibitem [{\citenamefont {Longhi}(2009)}]{Longhi2009PRL}%
  \BibitemOpen
  \bibfield  {author} {\bibinfo {author} {\bibfnamefont {S.}~\bibnamefont
  {Longhi}},\ }\href {https://doi.org/10.1103/PhysRevLett.103.123601}
  {\bibfield  {journal} {\bibinfo  {journal} {Phys. Rev. Lett.}\ }\textbf
  {\bibinfo {volume} {103}},\ \bibinfo {pages} {123601} (\bibinfo {year}
  {2009})}\BibitemShut {NoStop}%
\bibitem [{\citenamefont {Makris}\ \emph {et~al.}(2010)\citenamefont {Makris},
  \citenamefont {El-Ganainy}, \citenamefont {Christodoulides},\ and\
  \citenamefont {Musslimani}}]{Makris2010}%
  \BibitemOpen
  \bibfield  {author} {\bibinfo {author} {\bibfnamefont {K.~G.}\ \bibnamefont
  {Makris}}, \bibinfo {author} {\bibfnamefont {R.}~\bibnamefont {El-Ganainy}},
  \bibinfo {author} {\bibfnamefont {D.~N.}\ \bibnamefont {Christodoulides}},\
  and\ \bibinfo {author} {\bibfnamefont {Z.~H.}\ \bibnamefont {Musslimani}},\
  }\href {https://doi.org/10.1103/PhysRevA.81.063807} {\bibfield  {journal}
  {\bibinfo  {journal} {Phys. Rev. A}\ }\textbf {\bibinfo {volume} {81}},\
  \bibinfo {pages} {063807} (\bibinfo {year} {2010})}\BibitemShut {NoStop}%
\bibitem [{\citenamefont {Graefe}\ and\ \citenamefont
  {Jones}(2011)}]{GraefeJones2011PRA}%
  \BibitemOpen
  \bibfield  {author} {\bibinfo {author} {\bibfnamefont {E.-M.}\ \bibnamefont
  {Graefe}}\ and\ \bibinfo {author} {\bibfnamefont {H.~F.}\ \bibnamefont
  {Jones}},\ }\href {https://doi.org/10.1103/PhysRevA.84.013818} {\bibfield
  {journal} {\bibinfo  {journal} {Phys. Rev. A}\ }\textbf {\bibinfo {volume}
  {84}},\ \bibinfo {pages} {013818} (\bibinfo {year} {2011})}\BibitemShut
  {NoStop}%
\bibitem [{\citenamefont {Jones}(2014)}]{Jones2014}%
  \BibitemOpen
  \bibfield  {author} {\bibinfo {author} {\bibfnamefont {H.~F.}\ \bibnamefont
  {Jones}},\ }\href {https://doi.org/10.1007/s10773-014-2432-y} {\bibfield
  {journal} {\bibinfo  {journal} {Int. J. Theor. Phys.}\ }\textbf {\bibinfo
  {volume} {54}},\ \bibinfo {pages} {3986} (\bibinfo {year}
  {2014})}\BibitemShut {NoStop}%
\bibitem [{\citenamefont {Abdullaev}\ \emph {et~al.}(2010)\citenamefont
  {Abdullaev}, \citenamefont {Konotop}, \citenamefont {Salerno},\ and\
  \citenamefont {Yulin}}]{AbdullaevYulin2010PRE}%
  \BibitemOpen
  \bibfield  {author} {\bibinfo {author} {\bibfnamefont {F.~K.}\ \bibnamefont
  {Abdullaev}}, \bibinfo {author} {\bibfnamefont {V.~V.}\ \bibnamefont
  {Konotop}}, \bibinfo {author} {\bibfnamefont {M.}~\bibnamefont {Salerno}},\
  and\ \bibinfo {author} {\bibfnamefont {A.~V.}\ \bibnamefont {Yulin}},\ }\href
  {https://doi.org/10.1103/PhysRevE.82.056606} {\bibfield  {journal} {\bibinfo
  {journal} {Phys. Rev. E}\ }\textbf {\bibinfo {volume} {82}},\ \bibinfo
  {pages} {056606} (\bibinfo {year} {2010})}\BibitemShut {NoStop}%
\bibitem [{\citenamefont {Zhou}\ \emph {et~al.}(2010)\citenamefont {Zhou},
  \citenamefont {Guo}, \citenamefont {Wang},\ and\ \citenamefont
  {Liu}}]{ZhouLiu2010OL}%
  \BibitemOpen
  \bibfield  {author} {\bibinfo {author} {\bibfnamefont {K.}~\bibnamefont
  {Zhou}}, \bibinfo {author} {\bibfnamefont {Z.}~\bibnamefont {Guo}}, \bibinfo
  {author} {\bibfnamefont {J.}~\bibnamefont {Wang}},\ and\ \bibinfo {author}
  {\bibfnamefont {S.}~\bibnamefont {Liu}},\ }\href
  {https://doi.org/10.1364/OL.35.002928} {\bibfield  {journal} {\bibinfo
  {journal} {Opt. Lett.}\ }\textbf {\bibinfo {volume} {35}},\ \bibinfo {pages}
  {2928} (\bibinfo {year} {2010})}\BibitemShut {NoStop}%
\bibitem [{\citenamefont {Zhu}\ \emph {et~al.}(2011)\citenamefont {Zhu},
  \citenamefont {Wang}, \citenamefont {Zheng}, \citenamefont {Li},\ and\
  \citenamefont {He}}]{ZhuHe2011OL}%
  \BibitemOpen
  \bibfield  {author} {\bibinfo {author} {\bibfnamefont {X.}~\bibnamefont
  {Zhu}}, \bibinfo {author} {\bibfnamefont {H.}~\bibnamefont {Wang}}, \bibinfo
  {author} {\bibfnamefont {L.-X.}\ \bibnamefont {Zheng}}, \bibinfo {author}
  {\bibfnamefont {H.}~\bibnamefont {Li}},\ and\ \bibinfo {author}
  {\bibfnamefont {Y.-J.}\ \bibnamefont {He}},\ }\href
  {https://doi.org/10.1364/OL.36.002680} {\bibfield  {journal} {\bibinfo
  {journal} {Opt. Lett.}\ }\textbf {\bibinfo {volume} {36}},\ \bibinfo {pages}
  {2680} (\bibinfo {year} {2011})}\BibitemShut {NoStop}%
\bibitem [{\citenamefont {Nixon}\ \emph
  {et~al.}(2012{\natexlab{a}})\citenamefont {Nixon}, \citenamefont {Zhu},\ and\
  \citenamefont {Yang}}]{NixonYang2012OL}%
  \BibitemOpen
  \bibfield  {author} {\bibinfo {author} {\bibfnamefont {S.}~\bibnamefont
  {Nixon}}, \bibinfo {author} {\bibfnamefont {Y.}~\bibnamefont {Zhu}},\ and\
  \bibinfo {author} {\bibfnamefont {J.}~\bibnamefont {Yang}},\ }\href
  {https://doi.org/10.1364/OL.37.004874} {\bibfield  {journal} {\bibinfo
  {journal} {Opt. Lett.}\ }\textbf {\bibinfo {volume} {37}},\ \bibinfo {pages}
  {4874} (\bibinfo {year} {2012}{\natexlab{a}})}\BibitemShut {NoStop}%
\bibitem [{\citenamefont {Nixon}\ \emph
  {et~al.}(2012{\natexlab{b}})\citenamefont {Nixon}, \citenamefont {Ge},\ and\
  \citenamefont {Yang}}]{NixonYang2012PRA}%
  \BibitemOpen
  \bibfield  {author} {\bibinfo {author} {\bibfnamefont {S.}~\bibnamefont
  {Nixon}}, \bibinfo {author} {\bibfnamefont {L.}~\bibnamefont {Ge}},\ and\
  \bibinfo {author} {\bibfnamefont {J.}~\bibnamefont {Yang}},\ }\href
  {https://doi.org/10.1103/PhysRevA.85.023822} {\bibfield  {journal} {\bibinfo
  {journal} {Phys. Rev. A}\ }\textbf {\bibinfo {volume} {85}},\ \bibinfo
  {pages} {023822} (\bibinfo {year} {2012}{\natexlab{b}})}\BibitemShut
  {NoStop}%
\bibitem [{\citenamefont {Zhang}\ \emph {et~al.}(2021)\citenamefont {Zhang},
  \citenamefont {Chen}, \citenamefont {Wu}, \citenamefont {Busch},\ and\
  \citenamefont {Konotop}}]{ZhangKonotop2021PRL}%
  \BibitemOpen
  \bibfield  {author} {\bibinfo {author} {\bibfnamefont {Y.}~\bibnamefont
  {Zhang}}, \bibinfo {author} {\bibfnamefont {Z.}~\bibnamefont {Chen}},
  \bibinfo {author} {\bibfnamefont {B.}~\bibnamefont {Wu}}, \bibinfo {author}
  {\bibfnamefont {T.}~\bibnamefont {Busch}},\ and\ \bibinfo {author}
  {\bibfnamefont {V.~V.}\ \bibnamefont {Konotop}},\ }\href
  {https://doi.org/10.1103/physrevlett.127.034101} {\bibfield  {journal}
  {\bibinfo  {journal} {Phys. Rev. Lett.}\ }\textbf {\bibinfo {volume} {127}},\
  \bibinfo {pages} {034101} (\bibinfo {year} {2021})}\BibitemShut {NoStop}%
\bibitem [{\citenamefont {Salasnich}\ \emph {et~al.}(2002)\citenamefont
  {Salasnich}, \citenamefont {Parola},\ and\ \citenamefont
  {Reatto}}]{SalasnichReatto2002PRA}%
  \BibitemOpen
  \bibfield  {author} {\bibinfo {author} {\bibfnamefont {L.}~\bibnamefont
  {Salasnich}}, \bibinfo {author} {\bibfnamefont {A.}~\bibnamefont {Parola}},\
  and\ \bibinfo {author} {\bibfnamefont {L.}~\bibnamefont {Reatto}},\ }\href
  {https://doi.org/10.1103/PhysRevA.65.043614} {\bibfield  {journal} {\bibinfo
  {journal} {Phys. Rev. A}\ }\textbf {\bibinfo {volume} {65}},\ \bibinfo
  {pages} {043614} (\bibinfo {year} {2002})}\BibitemShut {NoStop}%
\bibitem [{\citenamefont {Lieb}\ \emph {et~al.}(2003)\citenamefont {Lieb},
  \citenamefont {Seiringer},\ and\ \citenamefont
  {Yngvason}}]{LiebSeiringer2003PRL}%
  \BibitemOpen
  \bibfield  {author} {\bibinfo {author} {\bibfnamefont {E.~H.}\ \bibnamefont
  {Lieb}}, \bibinfo {author} {\bibfnamefont {R.}~\bibnamefont {Seiringer}},\
  and\ \bibinfo {author} {\bibfnamefont {J.}~\bibnamefont {Yngvason}},\ }\href
  {https://doi.org/10.1103/PhysRevLett.91.150401} {\bibfield  {journal}
  {\bibinfo  {journal} {Phys. Rev. Lett.}\ }\textbf {\bibinfo {volume} {91}},\
  \bibinfo {pages} {150401} (\bibinfo {year} {2003})}\BibitemShut {NoStop}%
\bibitem [{\citenamefont {Pethick}\ and\ \citenamefont
  {Smith}(2008)}]{Pethick2008PethickCU}%
  \BibitemOpen
  \bibfield  {author} {\bibinfo {author} {\bibfnamefont {C.~J.}\ \bibnamefont
  {Pethick}}\ and\ \bibinfo {author} {\bibfnamefont {H.}~\bibnamefont
  {Smith}},\ }\href@noop {} {\emph {\bibinfo {title} {Bose--Einstein
  condensation in dilute gases}}}\ (\bibinfo  {publisher} {Cambridge university
  press},\ \bibinfo {year} {2008})\BibitemShut {NoStop}%
\bibitem [{\citenamefont {Haag}\ \emph {et~al.}(2014)\citenamefont {Haag},
  \citenamefont {Dast}, \citenamefont {L\"ohle}, \citenamefont {Cartarius},
  \citenamefont {Main},\ and\ \citenamefont {Wunner}}]{HaagWunner2014PRA}%
  \BibitemOpen
  \bibfield  {author} {\bibinfo {author} {\bibfnamefont {D.}~\bibnamefont
  {Haag}}, \bibinfo {author} {\bibfnamefont {D.}~\bibnamefont {Dast}}, \bibinfo
  {author} {\bibfnamefont {A.}~\bibnamefont {L\"ohle}}, \bibinfo {author}
  {\bibfnamefont {H.}~\bibnamefont {Cartarius}}, \bibinfo {author}
  {\bibfnamefont {J.}~\bibnamefont {Main}},\ and\ \bibinfo {author}
  {\bibfnamefont {G.}~\bibnamefont {Wunner}},\ }\href
  {https://doi.org/10.1103/PhysRevA.89.023601} {\bibfield  {journal} {\bibinfo
  {journal} {Phys. Rev. A}\ }\textbf {\bibinfo {volume} {89}},\ \bibinfo
  {pages} {023601} (\bibinfo {year} {2014})}\BibitemShut {NoStop}%
\bibitem [{\citenamefont {Gutöhrlein}\ \emph {et~al.}(2015)\citenamefont
  {Gutöhrlein}, \citenamefont {Schnabel}, \citenamefont {Iskandarov},
  \citenamefont {Cartarius}, \citenamefont {Main},\ and\ \citenamefont
  {Wunner}}]{GutohrleinWunner2015JoPA}%
  \BibitemOpen
  \bibfield  {author} {\bibinfo {author} {\bibfnamefont {R.}~\bibnamefont
  {Gutöhrlein}}, \bibinfo {author} {\bibfnamefont {J.}~\bibnamefont
  {Schnabel}}, \bibinfo {author} {\bibfnamefont {I.}~\bibnamefont
  {Iskandarov}}, \bibinfo {author} {\bibfnamefont {H.}~\bibnamefont
  {Cartarius}}, \bibinfo {author} {\bibfnamefont {J.}~\bibnamefont {Main}},\
  and\ \bibinfo {author} {\bibfnamefont {G.}~\bibnamefont {Wunner}},\ }\href
  {https://doi.org/10.1088/1751-8113/48/33/335302} {\bibfield  {journal}
  {\bibinfo  {journal} {J. Phys. A: Math. Theor.}\ }\textbf {\bibinfo {volume}
  {48}},\ \bibinfo {pages} {335302} (\bibinfo {year} {2015})}\BibitemShut
  {NoStop}%
\bibitem [{\citenamefont {Lye}\ \emph {et~al.}(2007)\citenamefont {Lye},
  \citenamefont {Fallani}, \citenamefont {Fort}, \citenamefont {Guarrera},
  \citenamefont {Modugno}, \citenamefont {Wiersma},\ and\ \citenamefont
  {Inguscio}}]{LyeInguscio2007PRA}%
  \BibitemOpen
  \bibfield  {author} {\bibinfo {author} {\bibfnamefont {J.~E.}\ \bibnamefont
  {Lye}}, \bibinfo {author} {\bibfnamefont {L.}~\bibnamefont {Fallani}},
  \bibinfo {author} {\bibfnamefont {C.}~\bibnamefont {Fort}}, \bibinfo {author}
  {\bibfnamefont {V.}~\bibnamefont {Guarrera}}, \bibinfo {author}
  {\bibfnamefont {M.}~\bibnamefont {Modugno}}, \bibinfo {author} {\bibfnamefont
  {D.~S.}\ \bibnamefont {Wiersma}},\ and\ \bibinfo {author} {\bibfnamefont
  {M.}~\bibnamefont {Inguscio}},\ }\href
  {https://doi.org/10.1103/PhysRevA.75.061603} {\bibfield  {journal} {\bibinfo
  {journal} {Phys. Rev. A}\ }\textbf {\bibinfo {volume} {75}},\ \bibinfo
  {pages} {061603} (\bibinfo {year} {2007})}\BibitemShut {NoStop}%
\bibitem [{\citenamefont {Biddle}\ and\ \citenamefont
  {Das~Sarma}(2010)}]{Biddle2010Biddle}%
  \BibitemOpen
  \bibfield  {author} {\bibinfo {author} {\bibfnamefont {J.}~\bibnamefont
  {Biddle}}\ and\ \bibinfo {author} {\bibfnamefont {S.}~\bibnamefont
  {Das~Sarma}},\ }\href {https://doi.org/10.1103/PhysRevLett.104.070601}
  {\bibfield  {journal} {\bibinfo  {journal} {Phys. Rev. Lett.}\ }\textbf
  {\bibinfo {volume} {104}},\ \bibinfo {pages} {070601} (\bibinfo {year}
  {2010})}\BibitemShut {NoStop}%
\bibitem [{\citenamefont {Zezyulin}\ and\ \citenamefont
  {Alfimov}(2024)}]{ZezyulinGeorgy2024PRA}%
  \BibitemOpen
  \bibfield  {author} {\bibinfo {author} {\bibfnamefont {D.~A.}\ \bibnamefont
  {Zezyulin}}\ and\ \bibinfo {author} {\bibfnamefont {G.~L.}\ \bibnamefont
  {Alfimov}},\ }\href {https://doi.org/10.1103/PhysRevA.110.063304} {\bibfield
  {journal} {\bibinfo  {journal} {Phys. Rev. A}\ }\textbf {\bibinfo {volume}
  {110}},\ \bibinfo {pages} {063304} (\bibinfo {year} {2024})}\BibitemShut
  {NoStop}%
\bibitem [{\citenamefont {Machholm}\ \emph {et~al.}(2004)\citenamefont
  {Machholm}, \citenamefont {Nicolin}, \citenamefont {Pethick},\ and\
  \citenamefont {Smith}}]{MachholmSmith2004PRAdoubling}%
  \BibitemOpen
  \bibfield  {author} {\bibinfo {author} {\bibfnamefont {M.}~\bibnamefont
  {Machholm}}, \bibinfo {author} {\bibfnamefont {A.}~\bibnamefont {Nicolin}},
  \bibinfo {author} {\bibfnamefont {C.~J.}\ \bibnamefont {Pethick}},\ and\
  \bibinfo {author} {\bibfnamefont {H.}~\bibnamefont {Smith}},\ }\href
  {https://doi.org/10.1103/PhysRevA.69.043604} {\bibfield  {journal} {\bibinfo
  {journal} {Phys. Rev. A}\ }\textbf {\bibinfo {volume} {69}},\ \bibinfo
  {pages} {043604} (\bibinfo {year} {2004})}\BibitemShut {NoStop}%
\bibitem [{\citenamefont {Zhang}\ and\ \citenamefont
  {Wu}(2009)}]{ZhangWu2009PRL}%
  \BibitemOpen
  \bibfield  {author} {\bibinfo {author} {\bibfnamefont {Y.}~\bibnamefont
  {Zhang}}\ and\ \bibinfo {author} {\bibfnamefont {B.}~\bibnamefont {Wu}},\
  }\href {https://doi.org/10.1103/PhysRevLett.102.093905} {\bibfield  {journal}
  {\bibinfo  {journal} {Phys. Rev. Lett.}\ }\textbf {\bibinfo {volume} {102}},\
  \bibinfo {pages} {093905} (\bibinfo {year} {2009})}\BibitemShut {NoStop}%
\bibitem [{\citenamefont {Dizdarevic}\ \emph {et~al.}(2015)\citenamefont
  {Dizdarevic}, \citenamefont {Dast}, \citenamefont {Haag}, \citenamefont
  {Main}, \citenamefont {Cartarius},\ and\ \citenamefont
  {Wunner}}]{DizdarevicWunner2015PRA}%
  \BibitemOpen
  \bibfield  {author} {\bibinfo {author} {\bibfnamefont {D.}~\bibnamefont
  {Dizdarevic}}, \bibinfo {author} {\bibfnamefont {D.}~\bibnamefont {Dast}},
  \bibinfo {author} {\bibfnamefont {D.}~\bibnamefont {Haag}}, \bibinfo {author}
  {\bibfnamefont {J.}~\bibnamefont {Main}}, \bibinfo {author} {\bibfnamefont
  {H.}~\bibnamefont {Cartarius}},\ and\ \bibinfo {author} {\bibfnamefont
  {G.}~\bibnamefont {Wunner}},\ }\href
  {https://doi.org/10.1103/PhysRevA.91.033636} {\bibfield  {journal} {\bibinfo
  {journal} {Phys. Rev. A}\ }\textbf {\bibinfo {volume} {91}},\ \bibinfo
  {pages} {033636} (\bibinfo {year} {2015})}\BibitemShut {NoStop}%
\bibitem [{\citenamefont {Pelinovsky}\ \emph {et~al.}(2013)\citenamefont
  {Pelinovsky}, \citenamefont {Kevrekidis},\ and\ \citenamefont
  {Frantzeskakis}}]{PelinovskyFrantzeskakis2013EL}%
  \BibitemOpen
  \bibfield  {author} {\bibinfo {author} {\bibfnamefont {D.~E.}\ \bibnamefont
  {Pelinovsky}}, \bibinfo {author} {\bibfnamefont {P.~G.}\ \bibnamefont
  {Kevrekidis}},\ and\ \bibinfo {author} {\bibfnamefont {D.~J.}\ \bibnamefont
  {Frantzeskakis}},\ }\href {https://doi.org/10.1209/0295-5075/101/11002}
  {\bibfield  {journal} {\bibinfo  {journal} {Europhysics Letters}\ }\textbf
  {\bibinfo {volume} {101}},\ \bibinfo {pages} {11002} (\bibinfo {year}
  {2013})}\BibitemShut {NoStop}%
\bibitem [{\citenamefont {Lin}\ \emph {et~al.}()\citenamefont {Lin},
  \citenamefont {Pi}, \citenamefont {Qi}, \citenamefont {Qin}, \citenamefont
  {Nori},\ and\ \citenamefont {Long}}]{Lin2025arxiv}%
  \BibitemOpen
  \bibfield  {author} {\bibinfo {author} {\bibfnamefont {H.}~\bibnamefont
  {Lin}}, \bibinfo {author} {\bibfnamefont {J.}~\bibnamefont {Pi}}, \bibinfo
  {author} {\bibfnamefont {Y.}~\bibnamefont {Qi}}, \bibinfo {author}
  {\bibfnamefont {W.}~\bibnamefont {Qin}}, \bibinfo {author} {\bibfnamefont
  {F.}~\bibnamefont {Nori}},\ and\ \bibinfo {author} {\bibfnamefont {G.-L.}\
  \bibnamefont {Long}},\ }\href {https://arxiv.org/abs/2404.16774} {\ }\Eprint
  {https://arxiv.org/abs/2404.16774} {arXiv:2404.16774} \BibitemShut {NoStop}%
\bibitem [{\citenamefont {Wu}\ and\ \citenamefont {Niu}(2003)}]{WuNiu2003NJP}%
  \BibitemOpen
  \bibfield  {author} {\bibinfo {author} {\bibfnamefont {B.}~\bibnamefont
  {Wu}}\ and\ \bibinfo {author} {\bibfnamefont {Q.}~\bibnamefont {Niu}},\
  }\href {https://doi.org/10.1088/1367-2630/5/1/104} {\bibfield  {journal}
  {\bibinfo  {journal} {New J. Phys.}\ }\textbf {\bibinfo {volume} {5}},\
  \bibinfo {pages} {104} (\bibinfo {year} {2003})}\BibitemShut {NoStop}%
\bibitem [{\citenamefont {Vitanov}\ and\ \citenamefont
  {Suominen}(1999)}]{VitanovSuominen1999PRA}%
  \BibitemOpen
  \bibfield  {author} {\bibinfo {author} {\bibfnamefont {N.~V.}\ \bibnamefont
  {Vitanov}}\ and\ \bibinfo {author} {\bibfnamefont {K.-A.}\ \bibnamefont
  {Suominen}},\ }\href {https://doi.org/10.1103/PhysRevA.59.4580} {\bibfield
  {journal} {\bibinfo  {journal} {Phys. Rev. A}\ }\textbf {\bibinfo {volume}
  {59}},\ \bibinfo {pages} {4580} (\bibinfo {year} {1999})}\BibitemShut
  {NoStop}%
\bibitem [{\citenamefont {Wang}\ \emph {et~al.}(2022)\citenamefont {Wang},
  \citenamefont {Sun},\ and\ \citenamefont {Liu}}]{WangJie2022PRA}%
  \BibitemOpen
  \bibfield  {author} {\bibinfo {author} {\bibfnamefont {W.-Y.}\ \bibnamefont
  {Wang}}, \bibinfo {author} {\bibfnamefont {B.}~\bibnamefont {Sun}},\ and\
  \bibinfo {author} {\bibfnamefont {J.}~\bibnamefont {Liu}},\ }\href
  {https://doi.org/10.1103/PhysRevA.106.063708} {\bibfield  {journal} {\bibinfo
   {journal} {Phys. Rev. A}\ }\textbf {\bibinfo {volume} {106}},\ \bibinfo
  {pages} {063708} (\bibinfo {year} {2022})}\BibitemShut {NoStop}%
\bibitem [{\citenamefont {Tiesinga}\ \emph {et~al.}(1993)\citenamefont
  {Tiesinga}, \citenamefont {Verhaar},\ and\ \citenamefont
  {Stoof}}]{TiesingaStoof1993PRA}%
  \BibitemOpen
  \bibfield  {author} {\bibinfo {author} {\bibfnamefont {E.}~\bibnamefont
  {Tiesinga}}, \bibinfo {author} {\bibfnamefont {B.~J.}\ \bibnamefont
  {Verhaar}},\ and\ \bibinfo {author} {\bibfnamefont {H.~T.~C.}\ \bibnamefont
  {Stoof}},\ }\href {https://doi.org/10.1103/PhysRevA.47.4114} {\bibfield
  {journal} {\bibinfo  {journal} {Phys. Rev. A}\ }\textbf {\bibinfo {volume}
  {47}},\ \bibinfo {pages} {4114} (\bibinfo {year} {1993})}\BibitemShut
  {NoStop}%
\bibitem [{\citenamefont {Inouye}\ \emph {et~al.}(1998)\citenamefont {Inouye},
  \citenamefont {Andrews}, \citenamefont {Stenger}, \citenamefont {Miesner},
  \citenamefont {Stamper-Kurn},\ and\ \citenamefont
  {Ketterle}}]{InouyeKetterle1998nature}%
  \BibitemOpen
  \bibfield  {author} {\bibinfo {author} {\bibfnamefont {S.}~\bibnamefont
  {Inouye}}, \bibinfo {author} {\bibfnamefont {M.~R.}\ \bibnamefont {Andrews}},
  \bibinfo {author} {\bibfnamefont {J.}~\bibnamefont {Stenger}}, \bibinfo
  {author} {\bibfnamefont {H.-J.}\ \bibnamefont {Miesner}}, \bibinfo {author}
  {\bibfnamefont {D.~M.}\ \bibnamefont {Stamper-Kurn}},\ and\ \bibinfo {author}
  {\bibfnamefont {W.}~\bibnamefont {Ketterle}},\ }\href
  {https://doi.org/10.1038/32354} {\bibfield  {journal} {\bibinfo  {journal}
  {Nature}\ }\textbf {\bibinfo {volume} {392}},\ \bibinfo {pages} {151}
  (\bibinfo {year} {1998})}\BibitemShut {NoStop}%
\bibitem [{\citenamefont {Sheng}\ \emph {et~al.}(2013)\citenamefont {Sheng},
  \citenamefont {Miri}, \citenamefont {Christodoulides},\ and\ \citenamefont
  {Xiao}}]{ShengXiao2013PRA}%
  \BibitemOpen
  \bibfield  {author} {\bibinfo {author} {\bibfnamefont {J.}~\bibnamefont
  {Sheng}}, \bibinfo {author} {\bibfnamefont {M.-A.}\ \bibnamefont {Miri}},
  \bibinfo {author} {\bibfnamefont {D.~N.}\ \bibnamefont {Christodoulides}},\
  and\ \bibinfo {author} {\bibfnamefont {M.}~\bibnamefont {Xiao}},\ }\href
  {https://doi.org/10.1103/PhysRevA.88.041803} {\bibfield  {journal} {\bibinfo
  {journal} {Phys. Rev. A}\ }\textbf {\bibinfo {volume} {88}},\ \bibinfo
  {pages} {041803} (\bibinfo {year} {2013})}\BibitemShut {NoStop}%
\bibitem [{\citenamefont {Wu}\ \emph {et~al.}(2014)\citenamefont {Wu},
  \citenamefont {Artoni},\ and\ \citenamefont {La~Rocca}}]{WuLaRocca2014PRL}%
  \BibitemOpen
  \bibfield  {author} {\bibinfo {author} {\bibfnamefont {J.-H.}\ \bibnamefont
  {Wu}}, \bibinfo {author} {\bibfnamefont {M.}~\bibnamefont {Artoni}},\ and\
  \bibinfo {author} {\bibfnamefont {G.~C.}\ \bibnamefont {La~Rocca}},\ }\href
  {https://doi.org/10.1103/PhysRevLett.113.123004} {\bibfield  {journal}
  {\bibinfo  {journal} {Phys. Rev. Lett.}\ }\textbf {\bibinfo {volume} {113}},\
  \bibinfo {pages} {123004} (\bibinfo {year} {2014})}\BibitemShut {NoStop}%
\bibitem [{\citenamefont {Zhang}\ \emph {et~al.}(2016)\citenamefont {Zhang},
  \citenamefont {Zhang}, \citenamefont {Sheng}, \citenamefont {Yang},
  \citenamefont {Miri}, \citenamefont {Christodoulides}, \citenamefont {He},
  \citenamefont {Zhang},\ and\ \citenamefont {Xiao}}]{ZhangXiao2016PRL}%
  \BibitemOpen
  \bibfield  {author} {\bibinfo {author} {\bibfnamefont {Z.}~\bibnamefont
  {Zhang}}, \bibinfo {author} {\bibfnamefont {Y.}~\bibnamefont {Zhang}},
  \bibinfo {author} {\bibfnamefont {J.}~\bibnamefont {Sheng}}, \bibinfo
  {author} {\bibfnamefont {L.}~\bibnamefont {Yang}}, \bibinfo {author}
  {\bibfnamefont {M.-A.}\ \bibnamefont {Miri}}, \bibinfo {author}
  {\bibfnamefont {D.~N.}\ \bibnamefont {Christodoulides}}, \bibinfo {author}
  {\bibfnamefont {B.}~\bibnamefont {He}}, \bibinfo {author} {\bibfnamefont
  {Y.}~\bibnamefont {Zhang}},\ and\ \bibinfo {author} {\bibfnamefont
  {M.}~\bibnamefont {Xiao}},\ }\href
  {https://doi.org/10.1103/PhysRevLett.117.123601} {\bibfield  {journal}
  {\bibinfo  {journal} {Phys. Rev. Lett.}\ }\textbf {\bibinfo {volume} {117}},\
  \bibinfo {pages} {123601} (\bibinfo {year} {2016})}\BibitemShut {NoStop}%
\bibitem [{\citenamefont {Tsuno}\ \emph {et~al.}(2026)\citenamefont {Tsuno},
  \citenamefont {Taie}, \citenamefont {Takasu}, \citenamefont {Yamashita},
  \citenamefont {Ozawa},\ and\ \citenamefont {Takahashi}}]{Tsuno2025Tsuno}%
  \BibitemOpen
  \bibfield  {author} {\bibinfo {author} {\bibfnamefont {T.}~\bibnamefont
  {Tsuno}}, \bibinfo {author} {\bibfnamefont {S.}~\bibnamefont {Taie}},
  \bibinfo {author} {\bibfnamefont {Y.}~\bibnamefont {Takasu}}, \bibinfo
  {author} {\bibfnamefont {K.}~\bibnamefont {Yamashita}}, \bibinfo {author}
  {\bibfnamefont {T.}~\bibnamefont {Ozawa}},\ and\ \bibinfo {author}
  {\bibfnamefont {Y.}~\bibnamefont {Takahashi}},\ }\href
  {https://doi.org/10.1038/s41467-025-67106-8} {\bibfield  {journal} {\bibinfo
  {journal} {Nat. Commun.}\ }\textbf {\bibinfo {volume} {17}} (\bibinfo {year}
  {2026})}\BibitemShut {NoStop}%
\bibitem [{\citenamefont {Li}\ \emph {et~al.}(2019)\citenamefont {Li},
  \citenamefont {Harter}, \citenamefont {Liu}, \citenamefont {de~Melo},
  \citenamefont {Joglekar},\ and\ \citenamefont {Luo}}]{LiLuo2019NatComm}%
  \BibitemOpen
  \bibfield  {author} {\bibinfo {author} {\bibfnamefont {J.}~\bibnamefont
  {Li}}, \bibinfo {author} {\bibfnamefont {A.~K.}\ \bibnamefont {Harter}},
  \bibinfo {author} {\bibfnamefont {J.}~\bibnamefont {Liu}}, \bibinfo {author}
  {\bibfnamefont {L.}~\bibnamefont {de~Melo}}, \bibinfo {author} {\bibfnamefont
  {Y.~N.}\ \bibnamefont {Joglekar}},\ and\ \bibinfo {author} {\bibfnamefont
  {L.}~\bibnamefont {Luo}},\ }\href
  {https://doi.org/10.1038/s41467-019-08596-1} {\bibfield  {journal} {\bibinfo
  {journal} {Nat. Commun.}\ }\textbf {\bibinfo {volume} {10}} (\bibinfo {year}
  {2019})}\BibitemShut {NoStop}%
\bibitem [{\citenamefont {Takasu}\ \emph {et~al.}(2020)\citenamefont {Takasu},
  \citenamefont {Yagami}, \citenamefont {Ashida}, \citenamefont {Hamazaki},
  \citenamefont {Kuno},\ and\ \citenamefont {Takahashi}}]{TakasuTakahashi2020}%
  \BibitemOpen
  \bibfield  {author} {\bibinfo {author} {\bibfnamefont {Y.}~\bibnamefont
  {Takasu}}, \bibinfo {author} {\bibfnamefont {T.}~\bibnamefont {Yagami}},
  \bibinfo {author} {\bibfnamefont {Y.}~\bibnamefont {Ashida}}, \bibinfo
  {author} {\bibfnamefont {R.}~\bibnamefont {Hamazaki}}, \bibinfo {author}
  {\bibfnamefont {Y.}~\bibnamefont {Kuno}},\ and\ \bibinfo {author}
  {\bibfnamefont {Y.}~\bibnamefont {Takahashi}},\ }\href
  {https://doi.org/10.1093/ptep/ptaa094} {\bibfield  {journal} {\bibinfo
  {journal} {Prog. Theor. Exp.}\ }\textbf {\bibinfo {volume} {2020}},\ \bibinfo
  {pages} {12A110} (\bibinfo {year} {2020})}\BibitemShut {NoStop}%
\bibitem [{\citenamefont {Ren}\ \emph {et~al.}(2022)\citenamefont {Ren},
  \citenamefont {Liu}, \citenamefont {Zhao}, \citenamefont {He}, \citenamefont
  {Pak}, \citenamefont {Li},\ and\ \citenamefont {Jo}}]{RenJo2022NaturePhys}%
  \BibitemOpen
  \bibfield  {author} {\bibinfo {author} {\bibfnamefont {Z.}~\bibnamefont
  {Ren}}, \bibinfo {author} {\bibfnamefont {D.}~\bibnamefont {Liu}}, \bibinfo
  {author} {\bibfnamefont {E.}~\bibnamefont {Zhao}}, \bibinfo {author}
  {\bibfnamefont {C.}~\bibnamefont {He}}, \bibinfo {author} {\bibfnamefont
  {K.~K.}\ \bibnamefont {Pak}}, \bibinfo {author} {\bibfnamefont
  {J.}~\bibnamefont {Li}},\ and\ \bibinfo {author} {\bibfnamefont {G.-B.}\
  \bibnamefont {Jo}},\ }\href {https://doi.org/10.1038/s41567-021-01491-x}
  {\bibfield  {journal} {\bibinfo  {journal} {Nat. Phys.}\ }\textbf {\bibinfo
  {volume} {18}},\ \bibinfo {pages} {385} (\bibinfo {year} {2022})}\BibitemShut
  {NoStop}%
\bibitem [{\citenamefont {Wang}\ \emph {et~al.}(2024)\citenamefont {Wang},
  \citenamefont {Li}, \citenamefont {Xie}, \citenamefont {Ding}, \citenamefont
  {Ji}, \citenamefont {Xiao}, \citenamefont {Jia}, \citenamefont {Yan},
  \citenamefont {Hu},\ and\ \citenamefont {Zhao}}]{WangZhao2024PRL}%
  \BibitemOpen
  \bibfield  {author} {\bibinfo {author} {\bibfnamefont {C.}~\bibnamefont
  {Wang}}, \bibinfo {author} {\bibfnamefont {N.}~\bibnamefont {Li}}, \bibinfo
  {author} {\bibfnamefont {J.}~\bibnamefont {Xie}}, \bibinfo {author}
  {\bibfnamefont {C.}~\bibnamefont {Ding}}, \bibinfo {author} {\bibfnamefont
  {Z.}~\bibnamefont {Ji}}, \bibinfo {author} {\bibfnamefont {L.}~\bibnamefont
  {Xiao}}, \bibinfo {author} {\bibfnamefont {S.}~\bibnamefont {Jia}}, \bibinfo
  {author} {\bibfnamefont {B.}~\bibnamefont {Yan}}, \bibinfo {author}
  {\bibfnamefont {Y.}~\bibnamefont {Hu}},\ and\ \bibinfo {author}
  {\bibfnamefont {Y.}~\bibnamefont {Zhao}},\ }\href
  {https://doi.org/10.1103/PhysRevLett.132.253401} {\bibfield  {journal}
  {\bibinfo  {journal} {Phys. Rev. Lett.}\ }\textbf {\bibinfo {volume} {132}},\
  \bibinfo {pages} {253401} (\bibinfo {year} {2024})}\BibitemShut {NoStop}%
\bibitem [{\citenamefont {Cheng}\ \emph {et~al.}(2026)\citenamefont {Cheng},
  \citenamefont {Tan}, \citenamefont {Xu},\ and\ \citenamefont
  {Lang}}]{Cheng2026ChengData}%
  \BibitemOpen
  \bibfield  {author} {\bibinfo {author} {\bibfnamefont {E.}~\bibnamefont
  {Cheng}}, \bibinfo {author} {\bibfnamefont {Y.}~\bibnamefont {Tan}}, \bibinfo
  {author} {\bibfnamefont {Y.}~\bibnamefont {Xu}},\ and\ \bibinfo {author}
  {\bibfnamefont {L.-J.}\ \bibnamefont {Lang}}\ }\href
  {https://doi.org/10.5281/zenodo.22029415} {10.5281/zenodo.22029415} (\bibinfo
  {year} {2026})\BibitemShut {NoStop}%
\bibitem [{\citenamefont {Gottlob}\ and\ \citenamefont
  {Schneider}(2023)}]{GottlobSchneider2023PRB}%
  \BibitemOpen
  \bibfield  {author} {\bibinfo {author} {\bibfnamefont {E.}~\bibnamefont
  {Gottlob}}\ and\ \bibinfo {author} {\bibfnamefont {U.}~\bibnamefont
  {Schneider}},\ }\href {https://doi.org/10.1103/PhysRevB.107.144202}
  {\bibfield  {journal} {\bibinfo  {journal} {Phys. Rev. B}\ }\textbf {\bibinfo
  {volume} {107}},\ \bibinfo {pages} {144202} (\bibinfo {year}
  {2023})}\BibitemShut {NoStop}%
\bibitem [{\citenamefont {Johnstone}\ \emph {et~al.}(2025)\citenamefont
  {Johnstone}, \citenamefont {Mishra}, \citenamefont {Zhu},\ and\ \citenamefont
  {Sanchez-Palencia}}]{JohnstoneSanchez2025PRA}%
  \BibitemOpen
  \bibfield  {author} {\bibinfo {author} {\bibfnamefont {D.}~\bibnamefont
  {Johnstone}}, \bibinfo {author} {\bibfnamefont {S.}~\bibnamefont {Mishra}},
  \bibinfo {author} {\bibfnamefont {Z.}~\bibnamefont {Zhu}},\ and\ \bibinfo
  {author} {\bibfnamefont {L.}~\bibnamefont {Sanchez-Palencia}},\ }\href
  {https://doi.org/10.1103/PhysRevA.111.043305} {\bibfield  {journal} {\bibinfo
   {journal} {Phys. Rev. A}\ }\textbf {\bibinfo {volume} {111}},\ \bibinfo
  {pages} {043305} (\bibinfo {year} {2025})}\BibitemShut {NoStop}%
\bibitem [{\citenamefont {Marzari}\ \emph {et~al.}(2012)\citenamefont
  {Marzari}, \citenamefont {Mostofi}, \citenamefont {Yates}, \citenamefont
  {Souza},\ and\ \citenamefont {Vanderbilt}}]{Marzari2012Marzari}%
  \BibitemOpen
  \bibfield  {author} {\bibinfo {author} {\bibfnamefont {N.}~\bibnamefont
  {Marzari}}, \bibinfo {author} {\bibfnamefont {A.~A.}\ \bibnamefont
  {Mostofi}}, \bibinfo {author} {\bibfnamefont {J.~R.}\ \bibnamefont {Yates}},
  \bibinfo {author} {\bibfnamefont {I.}~\bibnamefont {Souza}},\ and\ \bibinfo
  {author} {\bibfnamefont {D.}~\bibnamefont {Vanderbilt}},\ }\href
  {https://doi.org/10.1103/RevModPhys.84.1419} {\bibfield  {journal} {\bibinfo
  {journal} {Rev. Mod. Phys.}\ }\textbf {\bibinfo {volume} {84}},\ \bibinfo
  {pages} {1419} (\bibinfo {year} {2012})}\BibitemShut {NoStop}%
\bibitem [{\citenamefont {Peierls}(1933)}]{Peierls1933Peierls}%
  \BibitemOpen
  \bibfield  {author} {\bibinfo {author} {\bibfnamefont {R.}~\bibnamefont
  {Peierls}},\ }\href {https://doi.org/10.1007/bf01342591} {\bibfield
  {journal} {\bibinfo  {journal} {Zeitschrift f\"ur Physik}\ }\textbf {\bibinfo
  {volume} {80}},\ \bibinfo {pages} {763} (\bibinfo {year} {1933})}\BibitemShut
  {NoStop}%
\bibitem [{\citenamefont {Lang}\ \emph {et~al.}(2018)\citenamefont {Lang},
  \citenamefont {Wang}, \citenamefont {Wang},\ and\ \citenamefont
  {Chong}}]{LangChong2018PRB}%
  \BibitemOpen
  \bibfield  {author} {\bibinfo {author} {\bibfnamefont {L.-J.}\ \bibnamefont
  {Lang}}, \bibinfo {author} {\bibfnamefont {Y.}~\bibnamefont {Wang}}, \bibinfo
  {author} {\bibfnamefont {H.}~\bibnamefont {Wang}},\ and\ \bibinfo {author}
  {\bibfnamefont {Y.~D.}\ \bibnamefont {Chong}},\ }\href
  {https://doi.org/10.1103/PhysRevB.98.094307} {\bibfield  {journal} {\bibinfo
  {journal} {Phys. Rev. B}\ }\textbf {\bibinfo {volume} {98}},\ \bibinfo
  {pages} {094307} (\bibinfo {year} {2018})}\BibitemShut {NoStop}%
\bibitem [{\citenamefont {Hang}\ \emph {et~al.}(2013)\citenamefont {Hang},
  \citenamefont {Huang},\ and\ \citenamefont {Konotop}}]{HangVladimir2013PRL}%
  \BibitemOpen
  \bibfield  {author} {\bibinfo {author} {\bibfnamefont {C.}~\bibnamefont
  {Hang}}, \bibinfo {author} {\bibfnamefont {G.}~\bibnamefont {Huang}},\ and\
  \bibinfo {author} {\bibfnamefont {V.~V.}\ \bibnamefont {Konotop}},\ }\href
  {https://doi.org/10.1103/PhysRevLett.110.083604} {\bibfield  {journal}
  {\bibinfo  {journal} {Phys. Rev. Lett.}\ }\textbf {\bibinfo {volume} {110}},\
  \bibinfo {pages} {083604} (\bibinfo {year} {2013})}\BibitemShut {NoStop}%
\bibitem [{\citenamefont {Jin}(2017)}]{Jin2017PRA}%
  \BibitemOpen
  \bibfield  {author} {\bibinfo {author} {\bibfnamefont {L.}~\bibnamefont
  {Jin}},\ }\href {https://doi.org/10.1103/PhysRevA.96.032103} {\bibfield
  {journal} {\bibinfo  {journal} {Phys. Rev. A}\ }\textbf {\bibinfo {volume}
  {96}},\ \bibinfo {pages} {032103} (\bibinfo {year} {2017})}\BibitemShut
  {NoStop}%
\bibitem [{\citenamefont {Garmon}\ and\ \citenamefont
  {Noba}(2021)}]{GarmonNoba2021PRA}%
  \BibitemOpen
  \bibfield  {author} {\bibinfo {author} {\bibfnamefont {S.}~\bibnamefont
  {Garmon}}\ and\ \bibinfo {author} {\bibfnamefont {K.}~\bibnamefont {Noba}},\
  }\href {https://doi.org/10.1103/PhysRevA.104.062215} {\bibfield  {journal}
  {\bibinfo  {journal} {Phys. Rev. A}\ }\textbf {\bibinfo {volume} {104}},\
  \bibinfo {pages} {062215} (\bibinfo {year} {2021})}\BibitemShut {NoStop}%
\bibitem [{\citenamefont {Anastasiadis}\ \emph {et~al.}(2022)\citenamefont
  {Anastasiadis}, \citenamefont {Styliaris}, \citenamefont {Chaunsali},
  \citenamefont {Theocharis},\ and\ \citenamefont
  {Diakonos}}]{AnastasiadisFotios2022PRB}%
  \BibitemOpen
  \bibfield  {author} {\bibinfo {author} {\bibfnamefont {A.}~\bibnamefont
  {Anastasiadis}}, \bibinfo {author} {\bibfnamefont {G.}~\bibnamefont
  {Styliaris}}, \bibinfo {author} {\bibfnamefont {R.}~\bibnamefont
  {Chaunsali}}, \bibinfo {author} {\bibfnamefont {G.}~\bibnamefont
  {Theocharis}},\ and\ \bibinfo {author} {\bibfnamefont {F.~K.}\ \bibnamefont
  {Diakonos}},\ }\href {https://doi.org/10.1103/PhysRevB.106.085109} {\bibfield
   {journal} {\bibinfo  {journal} {Phys. Rev. B}\ }\textbf {\bibinfo {volume}
  {106}},\ \bibinfo {pages} {085109} (\bibinfo {year} {2022})}\BibitemShut
  {NoStop}%
\bibitem [{\citenamefont {Yin}\ \emph {et~al.}()\citenamefont {Yin},
  \citenamefont {Tang}, \citenamefont {Tan}, \citenamefont {Bakhat},
  \citenamefont {Dong}, \citenamefont {Zhu},\ and\ \citenamefont
  {Yang}}]{Yin2024arxiv}%
  \BibitemOpen
  \bibfield  {author} {\bibinfo {author} {\bibfnamefont {K.}~\bibnamefont
  {Yin}}, \bibinfo {author} {\bibfnamefont {K.}~\bibnamefont {Tang}}, \bibinfo
  {author} {\bibfnamefont {L.}~\bibnamefont {Tan}}, \bibinfo {author}
  {\bibfnamefont {S.~I.~D.}\ \bibnamefont {Bakhat}}, \bibinfo {author}
  {\bibfnamefont {T.}~\bibnamefont {Dong}}, \bibinfo {author} {\bibfnamefont
  {H.}~\bibnamefont {Zhu}},\ and\ \bibinfo {author} {\bibfnamefont
  {Y.}~\bibnamefont {Yang}},\ }\href {https://arxiv.org/abs/2411.00591} {\
  }\Eprint {https://arxiv.org/abs/2411.00591} {arXiv:2411.00591} \BibitemShut
  {NoStop}%
\bibitem [{\citenamefont {Gong}\ \emph {et~al.}(2018)\citenamefont {Gong},
  \citenamefont {Ashida}, \citenamefont {Kawabata}, \citenamefont {Takasan},
  \citenamefont {Higashikawa},\ and\ \citenamefont {Ueda}}]{Gong2018Gong}%
  \BibitemOpen
  \bibfield  {author} {\bibinfo {author} {\bibfnamefont {Z.}~\bibnamefont
  {Gong}}, \bibinfo {author} {\bibfnamefont {Y.}~\bibnamefont {Ashida}},
  \bibinfo {author} {\bibfnamefont {K.}~\bibnamefont {Kawabata}}, \bibinfo
  {author} {\bibfnamefont {K.}~\bibnamefont {Takasan}}, \bibinfo {author}
  {\bibfnamefont {S.}~\bibnamefont {Higashikawa}},\ and\ \bibinfo {author}
  {\bibfnamefont {M.}~\bibnamefont {Ueda}},\ }\href
  {https://doi.org/10.1103/PhysRevX.8.031079} {\bibfield  {journal} {\bibinfo
  {journal} {Phys. Rev. X}\ }\textbf {\bibinfo {volume} {8}},\ \bibinfo {pages}
  {031079} (\bibinfo {year} {2018})}\BibitemShut {NoStop}%
\end{thebibliography}%
	\bibliographystyle{apsrev4-2}
\end{document}